\documentclass[useAMS,usenatbib,usegraphicx]{mn2e}

\usepackage{amssymb}
\usepackage{fixltx2e}

\newcommand{\FAP}{{\rm FAP}}

\title[The riddle of HD~82943]{Possible solution to the riddle of HD~82943 multi-planet
system: the three-planet resonance 1:2:5?}

\author[R.V.~Baluev \& C.~Beaug{\'e}]{
Roman V. Baluev$^{1,2}$\thanks{E-mail: r.baluev@spbu.ru} and Cristian Beaug{\'e}$^{3}$\\
$^1$Central Astronomical Observatory at Pulkovo of Russian Academy of Sciences,
Pulkovskoje shosse 65, St Petersburg 196140, Russia\\
$^2$Sobolev Astronomical Institute, St Petersburg State University, Universitetskij
prospekt 28, Petrodvorets, St Petersburg 198504, Russia\\
$^3$Instituto de Astronom\'ia Te\'orica y Experimental, Observatorio Astron\'omico,
Universidad Nacional de C\'ordoba, \\ Laprida 854, (X5000BGR) C\'ordoba, Argentina}

\begin{document}

\date{Accepted 2013 December 27.
      Received 2013 December 13;
      in original form 2013 October 26}

\pagerange{\pageref{firstpage}--\pageref{lastpage}} \pubyear{2014}

\maketitle

\label{firstpage}

\begin{abstract}
We carry out a new analysis of the published radial velocity data for the planet-hosting
star HD82943. We include the recent Keck/HIRES measurements as well as the aged but much
more numerous CORALIE data. We find that the CORALIE radial velocity measurements are
polluted by a systematic annual variation which affected the robustness of many previous
results. We show that after purging this variation, the residuals still contain a clear
signature of an additional $\sim 1100$~days periodicity. The latter variation leaves
significant hints in all three independent radial velocity subsets that we analysed: the
CORALIE data, the Keck data acquired prior to a hardware upgrade, and the Keck data taken
after the upgrade.

We mainly treat this variation as a signature of a third planet in the system, although we
cannot rule out other interpretations, such as long-term stellar activity. We find it easy
to naturally obtain a stable three-planet radial-velocity fit close to the three-planet
mean-motion resonance 1:2:5, with the two main planets (those in the 1:2 resonance) in an
aligned apsidal corotation. The dynamical status of the third planet is still uncertain:
it may reside in as well as slightly out of the 5:2 resonance. We obtain the value of
about $1075$~days for its orbital period and $\sim 0.3 M_{\rm Jup}$ for its minimum mass,
while the eccentric parameters are uncertain.
\end{abstract}


\section{Introduction}
CORALIE radial velocity (RV) measurements \citep[e.g.][]{Mayor04} imply that the planetary
system of HD~82943 includes at least two giant planets moving in the $2:1$ mean-motion
resonance (MMR). However, although this $2:1$ MMR was identified long ago
\citep{GozdMac01}, there was no a consensus concerning the orbital parameters of the major
planets, or even what is the total number of the planets in the system. Only the periods
of these two planets were determined with more or less good precision: $P_c \approx 220$~d
and $P_b \approx 440$~d. The main uncertainty was related to the orbital eccentricities
and the associated periapses arguments. The only certain assertion was that the
eccentricity $e_c$ or both the eccentricities are large. In such a case the dynamical
regime of this system remains poorly constrained: the original CORALIE RV data allow a lot
of alternative orbital configurations, both stable or unstable, and without any clear
advantage in the goodness of the fit \citep{FerrazMello05}.

The addition of $23$ Keck measurements by \citet{Lee06} did not improve the situation very
much. It appeared that the data contain the hint of an extra planet with an uncertain
$P_d\sim 1000$~d \citep{GozdKon06}, and that such three-planet system may lie close to a
Laplace resonance with $P_c:P_b:P_d\approx 1:2:4$ \citep{Beauge08}. But including an extra
planet to the RV curve model makes the reliable fitting of the data even more difficult.
Besides, it follows from the analysis done by \citet{Beauge08} that there is some
suspicious discrepancy between the CORALIE and Keck data. The orbital fits were rather
sensitive to removing some individual RV measurements or a set of measurements. Another
alternative planetary configuration of HD~82943, involving the 1:1 MMR, was provided by
\citep{GozdKon06}.

Recently, \citet{Tan13} published a new analysis with a significantly expanded Keck data
set for HD~82943. They suggested a stable two-planet fit, corresponding to an aligned
apsidal corotation state. Some hints of the third planets with $P_d\sim 1000$~d were noted
by \citet{Tan13} too, but contrary to \citet{Beauge08}, this variation was found to have
insufficient statistical significance. We would like to highlight that the conclusions
drawn by \citet{Beauge08} about the third planet were based mainly on the CORALIE data,
while \citet{Tan13} did not use the CORALIE data at all, suspecting them unreliable. With
this in mind, a solely independent (even if marginal) detection of the $\sim 1000$~d
variation made by \citet{Tan13} should be considered as a further argument \emph{in
favour} of its existence. At least, it would be incorrect to say that the \citet{Tan13}
work retracts this variation.

\citet{Tan13} decision to not rely on the CORALIE data looks pretty justified at this
step. Indeed, numerous previous works made it rather obvious that CORALIE and Keck data
for HD~82943 refuse to play together. It seems likely that there is some extra RV
variation that contaminates one or even both these data sets, making them contradicting
with each other at some stage. The main goal of the present paper is to carry out a
self-consistent joint analysis of these two RV data sets. The CORALIE data set still
outnumbers the combined Keck one more than by the factor of $2$, so it is highly
undesirable to be disregarded. Besides, we would like to bring some clarity concerning the
putative third planet, since \citet{Tan13} in fact neither confirmed nor retracted it. We
do not consider here the 1:1 MMR solution introduced by \citet{GozdKon06}.

The paper is organised as follows. In Sect.~\ref{sec_data} we describe in more detail the
data that we use in our analysis. In Sect.~\ref{sec_start} we demonstrate that a plain
analysis of the merged RV data leads us to an unrealistic dynamically unstable orbital
configuration. In Sect.~\ref{sec_acr} we discuss the quick method of obtaining a
dynamically stable orbital fit that relies on the theory of apsidal corotation resonances.
In Sect.~\ref{sec_deep} we carry out an in-depth analysis of the combined RV data. In
particular, we try to identify the sources that make the Keck and CORALIE data
inconsistent with each other and also to assess the detectability of the putative third
planet in the combined RV time series. We give our final orbital fits for HD~82943 in that
section. In Sect.~\ref{sec_valid} we justify the robustness of the statistical analysis
methods that we used in the work. In Sect.~\ref{sec_third} we discuss all pro and contra
concerning the existence of the putative third planet. Sect.~\ref{sec_inc} is devoted to
the uncertainty of the system orbital inclination and its impact on the data analysis and
planetary dynamics. In Sect.~\ref{sec_dyn} we consider the dynamics of our three-planet
configurations, paying particular attention to the most uncertain orbital parameters. In
Sect.~\ref{sec_migr} we provide the simulations of the three-planet migration and discuss
the conditions leading to the capture in the three-planet resonance.

\section{Radial velocity data}
\label{sec_data}
We used several publicly available data sets in the work. First, there are $N=142$ CORALIE
measurements from \citet{Mayor04} with typical uncertainty $4-5$~m/s and the time span of
$\sim 4.4$~yr. These data were never released in a table form, and we scanned them out of
the relevant EPS figure available at the \texttt{arXiv.org} preprint of \citep{Mayor04}. A
similar procedure was applied by \citet{FerrazMello05,Lee06,GozdKon06,Beauge08}. All
coordinates are stored in the EPS file as integer values, so the extracted data should
inevitably contain some additional round-off errors. The maximum round-off error of the
restored time is $\pm 0.5$~d, and of the restored radial velocity is $\pm 0.07$~m/s. As we
believe, the RV uncertainties can be reconstructed from the figure without additional
errors, because it seems that they were already rounded to integer numbers by
\citet{Mayor04} before plotting the figure (the precision of $1$~m/s in the RV uncertainty
is typical for other public ELODIE/CORALIE data sets). We just ensured that reconstructed
RV uncertainties are all close to integer numbers and then rounded them to get rid of the
scanning errors. On contrary, the reconstructed RV measurements do not concentrate near
integer values, indicating that their initial precision was probably better than $1$~m/s
(maybe $0.1$~m/s). Thus we left them without any further postprocessing.

The expected distribution of the EPS scanning errors is the uniform one for the time and
the symmetric triangular one for the radial velocity (because we determined the RV value
as half-sum of its error bar limits). Therefore, the implied standard deviations of these
errors should be $\sigma_t \approx 0.5/\sqrt{3} \approx 0.3$~d for the time and
$\sigma_{RV} \approx 0.07/\sqrt{6} \approx 0.03$~m/s for the radial velocity.

Obviously, only the time errors represent a potential issue. Let us assume that the RV
curve is given by the model $\mu(t)$. Then we may say that the time uncertainty $\sigma_t$
acts as an indirectly induced random (non-Gaussian) RV noise generating extra RV
uncertainty of $\sim |\mu'(t)| \sigma_t$, where $\mu'(t)$ is the star radial acceleration
induced by all orbiting planets. From the likely orbital parameters we can limit this
acceleration by roughly $5$~m/s/d. The maximum is achieved when the massive planets pass
their pericenters simultaneously. Thus, the additional RV uncertainty indirectly induced
by $\sigma_t \approx 0.3$~d is $1.5$~m/s at worst. This is still well below the best
residual r.m.s. of the CORALIE data that we obtain in this paper, $\sim 7$~m/s. It is
unlikely that extra errors of $\sim 1$~m/s or so may introduce significant changes in the
RV curve parameters, given the primary noise component of $\sim 7$~m/s, and given the fact
that RV uncertainties were already rounded to $1$~m/s precision by \citet{Mayor04}. The
scanning errors may slightly increase the estimated values of the CORALIE RV jitter given
below, but in average this shift is properly taken into account by our fitting algorithm
\citep{Baluev08b}, and actually it appears quite negligible (recall that statistical
uncertainties sum via their squared values). Given this argumentation, we believe that it
is pretty safe to use our reconstructed CORALIE data in practice, unless shorter orbital
periods like $\sim 10$~d get involved.

In addition to CORALIE, there are recent Keck data available in \citet{Tan13}. According
to the recommendations by \citet{Tan13}, we split these data in two independent subsets,
before and after a hardware upgrade. The first Keck subset consists of only $N=22$
measurements with the average stated uncertainty of $\sim 1.5$~m/s and the time span of
$\sim 3.2$~yr. The second Keck subset contains $N=42$ measurements spanning $6.3$~yr and
having the typical stated uncertainties of $1-1.5$~m/s. The CORALIE and the first Keck
data set notably overlap with each other, while the second Keck data set does not overlap
with any of the others. Notice that although the CORALIE data are older and less accurate,
they outnumber the Keck data more than by the factor of $2$, and also expand the time
base. Therefore, the contribution of the CORALIE and the Keck data in the results of the
analysis should be roughly equal: none should be disregarded. However, as follows from the
analysis made by \citet{Beauge08}, there is some inconsistency between the Keck and
CORALIE data sets which makes their joint analysis unreliable. This forced \citet{Tan13}
to discard the CORALIE data from their analysis. One of the underlying goals of the
present paper is to find a way to merge these data sets in a consistent manner. The
combined time series contains $N=206$ data points covering $12.4$~yrs cumulatively.

\section{Preliminary data analysis}
\label{sec_start}
To carry out the RV data fitting, we use the maximum-likelihood method from
\citet{Baluev08b}, which was implemented in the PlanetPack software \citep{Baluev13c}.
This method allows to adaptively fit the RV curve parameters together with the parameters
of the RV noise (``jitter''). It is especially useful for analysing the mixed
heterogeneous time series, which we have here: different data sets are allowed to have
different values of the RV jitter (which in practice is a typical case), and this makes
them weighted in a considerably more adequate way. To compare different best fitting
models we will mainly rely on the adjusted likelihood-ratio statistic $\tilde Z$ from
\citep{Baluev08b,Baluev13c}, and assuming that $\tilde Z$ obeys a chi-square distribution
asymptotically (for $N\to\infty$). We verify the practical validity of this approach in
Sect.~\ref{sec_valid} below.

First of all, we tried to fit the available RV data for HD82943 with the Keplerian and the
Newtonian ($N$-body) two-planet models. Both fits correspond to an anti-aligned apsidal
configuration of the planets. This configuration appears highly unstable: the system does
not survive even $1000$~years of the dynamical simulation. Although stable configurations
with anti-aligned apses are possible \citep{Ferraz-Mello-lec1,Beauge03,Beauge06}, and even
the one was once reported for HD82943 \citep{Ji03}, the anti-aligned configuration that
follows from the \emph{present} data is highly unstable, mainly due to unsuitable values
of the eccentricites (to have a stable anti-aligned ACR, the eccentricities must be much
larger, implying intersecting orbits).

\begin{table}
\caption{Best fitting parameters of the HD82943 planetary system: two-planet Newtonian
edge-on model}
\label{tab_bc}
\begin{tabular}{llll}
\hline
\multicolumn{4}{c}{planetary orbital parameters and masses}  \\
                       & planet c         & planet b       & \\
$P$~[day]              & $220.078(51)$    & $441.47(35)$   & \\
$K$~[m/s]              & $65.4(1.5)$      & $41.91(77)$    & \\
$e$                    & $0.3663(97)$     & $0.162(36)$    & \\
$\omega$~[$^\circ$]    & $117.2(1.7)$     & $300.9(3.4)$   & \\
$\lambda$~[$^\circ$]   & $308.7(1.4)$     & $216.4(1.0)$   & \\
$M$~[$M_{\rm Jup}$]    & $1.959(47)$      & $1.681(28)$    & \\
$a$~[AU]               & $0.74345(12)$    & $1.18306(62)$  & \\
$i$~[$^\circ$]         &\multicolumn{2}{c}{$90 \rm (fixed)$} \\
\hline
\multicolumn{4}{c}{parameters of the data sets} \\
                           & CORALIE        & Keck 1          & Keck 2      \\
$c$~[m/s]                  & $8144.18(80)$  & $-6.1(1.9)$     & $-7.68(62)$ \\
$\sigma_{\rm jitter}$~[m/s]& $6.63(58)$     & $9.0(1.4)$      & $3.81(46)$  \\
r.m.s.~[m/s]               & $7.85$         & $8.83$          & $3.87$      \\
\hline
\multicolumn{4}{c}{general characteristics of the fit}  \\
$\tilde l$~[m/s]           & \multicolumn{3}{c}{$7.11$} \\
$d$                        & \multicolumn{3}{c}{$16$}   \\
\hline
\end{tabular}\\
The parameters have the following meaning: orbital period $P$, RV semiamplitude $K$,
eccentricity $e$, pericenter argument $\omega$, mean longitude $\lambda$. The planets mass
$M$ and the semimajor axis $a$ values were derived assuming the inclination of
$i=90^\circ$ and the mass of the star $M_\star = 1.13 M_\odot$, taken from \citet{Tan13}.
The uncertainty of $M_\star$ was not included in the uncertainties of the derived values.
The parameter $c$ is the constant RV offset, and $\sigma_{\rm jitter}$ is the estimated RV
jitter (individual for each data set). The fit epoch is $JD2453500$, and the elements are
in the Jacobi reference frame described in \citep{Baluev11}. The goodness of the fit
$\tilde l$ is tied to the modified likelihood function as explained in \citep{Baluev08b}.
The integer $d$ is the total number of free parameters.
\end{table}

We tried to fit edge-on as well as an inclined (though still coplanar) Newtonian model,
but this did not make any significant changes to the best fitting parameters. The
Newtonian edge-on fit is given in Table~\ref{tab_bc}. Although we obtained some meaningful
and apparently rather promising estimation of the orbital inclination of $44^\circ\pm
14^\circ$, from the likelihood-ratio test we find that it is statistically consistent with
the edge-on fit: the relevant statistical significance is only $1.3$-sigma. Actually, even
the Keplerian fit does not differ much from the Newtonian one. To compare such non-nested
models we can use the Vuong test \citep{Vuong89,Baluev12}, which yields only rather
marginal $1.8$-sigma separation between the Keplerian and the Newtonian edge-on
fit.\footnote{Basically, Vuong test is a modification of the likelihood ratio test
carrying a special normalization. The Voung statistic is equal to zero for models with
equal maximum likelihood and increases when the discrepancy between the models increases.}

Remarkably, the anti-aligned configuration in Table~\ref{tab_bc} is significantly
different from the aligned one obtained by \citet{Tan13}, which was stable. Obviously,
this change is the effect of the CORALIE data that push the best fitting configuration in
some unsuitable direction. The CORALIE data, or even both Keck and CORALIE data probably
contain some additional variations that we need to identify and eliminate before we may
obtain any reliable results. As the planetary configuration of Table~\ref{tab_bc} is
severely unstable, it is unrealistic and is shown here mainly for demonstrative purposes.
We should apply some more intricate data analysis method to obtain a more realistic
orbital fit based on the combined RV data.

\section{Stability and the value of the apsidal corotation resonances}
\label{sec_acr}
The most easy and direct way to identify any possible spurious variations in the data is
to investigate the RV residuals left after subtraction of the \emph{true} planetary RV
contributions. However, we do not have the true orbital configuration of the system at our
disposal; what we have is only an unrealistic unstable configuration distorted by the
spurious variation that we want to eradicate. The formal best fitting solution tries to
compensate this variation by means of some bias in the planetary parameters. The best fit
thus becomes unstable and, on the other hand, the polluting variation remains hidden in
the noise. To bring this variation to the light, we may force the orbital fit to be more
physically realistic. For example, we may require it to be dynamically stable. This would
bring the fit more close to the true configuration, while the polluting RV variation would
become more obvious. This would help us to identify it among other (irrelevant or noisy)
peaks of a periodogram.

So, how we can find a realistic stable two-planet configuration, if the RV data do not
reveal it to us immediately? There are a lot of works devoted to this issue. For example,
in \citet{GozdMac01,Gozd05,GozdKon06,Gozd08} it was suggested to penalize the RV
goodness-of-fit function with the MEGNO chaoticity indicator. Although this is a direct
and minimum-force approach, we do not use it here, because it looks too slow for our
goals. Besides, it acts as a very irregular constraint imposed on the orbital parameters,
and its irregularity disables any reliable statistical treatment. We need to assess the
statistical reliability of the results, and also to reveal the actual agent that makes the
best fit unstable. To fulfil these goals, we will use another approach, initially
suggested in \citep{Baluev08c}, which is based on the theory of Apsidal Corotation
Resonances (ACRs). A quick way to stabilise a high-eccentricity resonant planetary system
is to fix it in an exact ACR. This makes the resulting best fitting configuration surely
stable. The ACR constraint is excessive: the stablility does not necessarily requre ACR.
For example, low-eccentricity orbits are usually stable without any ACRs. But in the
particular case of HD82943 the ACR is a natural way to stabilize the system, because of
its high orbital eccentricities. Besides, the ACR configuration is likely close to the
truth: it follows e.g. from the results by \citet{Tan13}. There are also arguments related
to the planet migration that make the ACR assumption rather realistic and desirable.

The theory of the ACRs and their relation to the planetary migration is explained in
\citet{Beauge03,Ferraz-Mello-lec1,Beauge06,Michtchenko06}. The further justification of
this ACR fitting method, as well as technical details are given in \citet{Baluev08c}, and
an implementation is available in PlanetPack. In short, the ACR condition puts $4$
equality constraints on the entire system of orbital parameters and on the planetary mass
ratio. Note that for coplanar orbital configurations that we consider here, the ratio of
the planet masses $M_k$ is equal to the ratio of the relevant minimum masses $M_k \sin i$,
implying that the uncertain value of $i$ does not significantly affect the imposed ACR
constraint.

\begin{table}
\caption{Best fitting parameters of the HD82943 planetary system: two-planet Newtonian
edge-on ACR model}
\label{tab_bc_ACR}
\begin{tabular}{llll}
\hline
\multicolumn{4}{c}{planetary orbital parameters and masses} \\
                       & planet c         & planet b       & \\
$P$~[day]              & $220.067(42)$    & $439.611(94)$  & \\
$K$~[m/s]              & $53.18(69)$      & $41.31(70)$    & \\
$e$                    & $0.432(13)$      & $0.1468(68)$   & \\
$\omega$~[$^\circ$]    & $120.9(1.3)$     & $120.9(1.3)$   & \\
$\lambda$~[$^\circ$]   & $309.84(57)$     & $216.36(82)$   & \\
$M$~[$M_{\rm Jup}$]    & $1.545(23)$      & $1.658(28)$    & \\
$a$~[AU]               & $0.743338(94)$   & $1.17959(17)$  & \\
$i$~[$^\circ$]         &\multicolumn{2}{c}{$90 \rm (fixed)$} \\
\hline
\multicolumn{4}{c}{parameters of the data sets} \\
                           & CORALIE        & Keck 1          & Keck 2      \\
$c$~[m/s]                  & $8142.40(77)$  & $-5.5(1.7)$     & $-7.82(68)$ \\
$\sigma_{\rm jitter}$~[m/s]& $7.42(61)$     & $8.1(1.3)$      & $4.21(50)$  \\
r.m.s.~[m/s]               & $8.55$         & $8.00$          & $4.27$      \\
\hline
\multicolumn{4}{c}{general characteristics of the fit} \\
$\tilde l$~[m/s]           & \multicolumn{3}{c}{$7.55$} \\
$d$                        & \multicolumn{3}{c}{$16-4=12$}   \\
\hline
\end{tabular}\\
See notes of Table~\ref{tab_bc}. The number of the degrees of freedom is reduced by $4$
due to the ACR constraint imposed on the planets \emph{c} and \emph{b}.
\end{table}

In Table~\ref{tab_bc_ACR} we give an ACR version of the fit from Table~\ref{tab_bc}. As we
expect, this ACR configuration should be more close to the truth than the one in
Table~\ref{tab_bc}. The likelihood-ratio separation between the fits of
Tables~\ref{tab_bc} and~\ref{tab_bc_ACR} is very significant: about $4.2$-sigma in the
asymptotic approximation. To identify the source of this difference we need to investigate
the residuals of the both fits, and then to compare them with each other.

\section{In-depth data analysis}
\label{sec_deep}
\begin{figure}
\includegraphics[width=84mm]{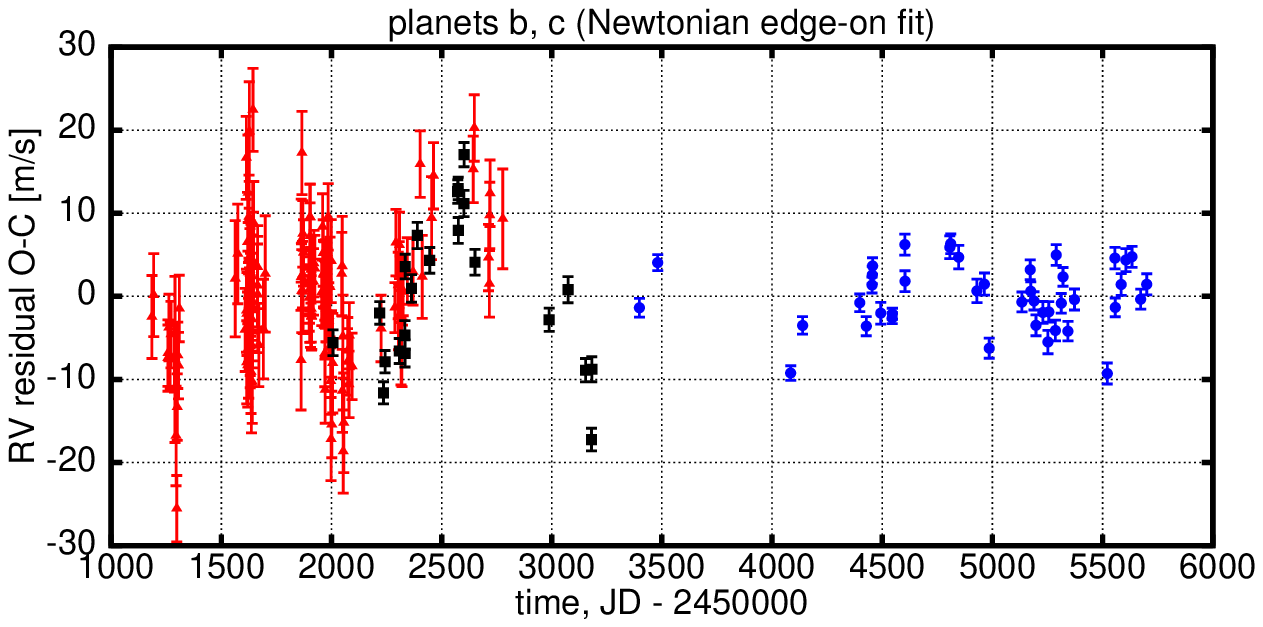}\\
\includegraphics[width=84mm]{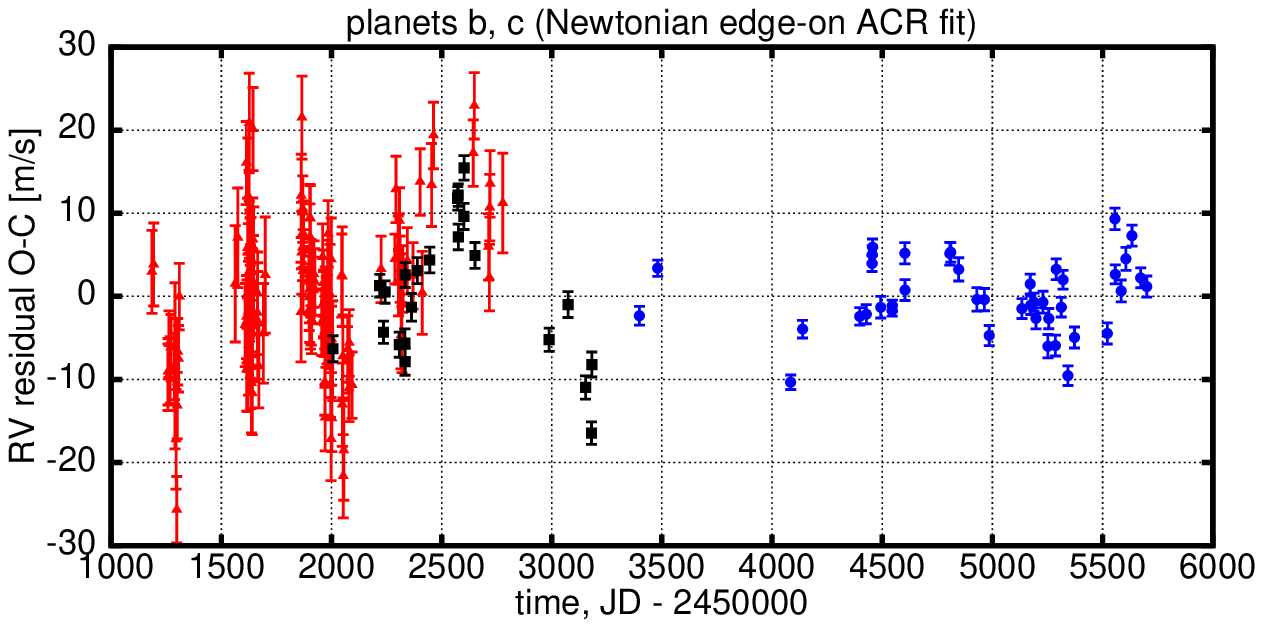}
\caption{The RV residuals to the best fitting two-planet models of HD82943. The CORALIE
data are marked as red triangles, the first Keck data-set by black squares, and the second
Keck data-set by blue circles. The error bars display only the internal RV uncertainties
(without the jitter).}
\label{fig_rsd}
\end{figure}

First of all, let us just plainly look at the RV residuals of the best-fitting models in
Fig.~\ref{fig_rsd}. Although the differential variation between these residuals is not yet
obvious, one thing is clear: these residuals are not consistent with a pure noise. For
example, the Keck-1 data show a systematic deviation which is partly confirmed by the
CORALIE data that overlap with this Keck range. The density of the CORALIE data does not
allow to see any further details inside their cloud, however. The presence of any residual
variation in the Keck-2 data is unclear, though some hints might be spotted in the ACR
case.

\begin{figure*}
\includegraphics[width=0.95\linewidth]{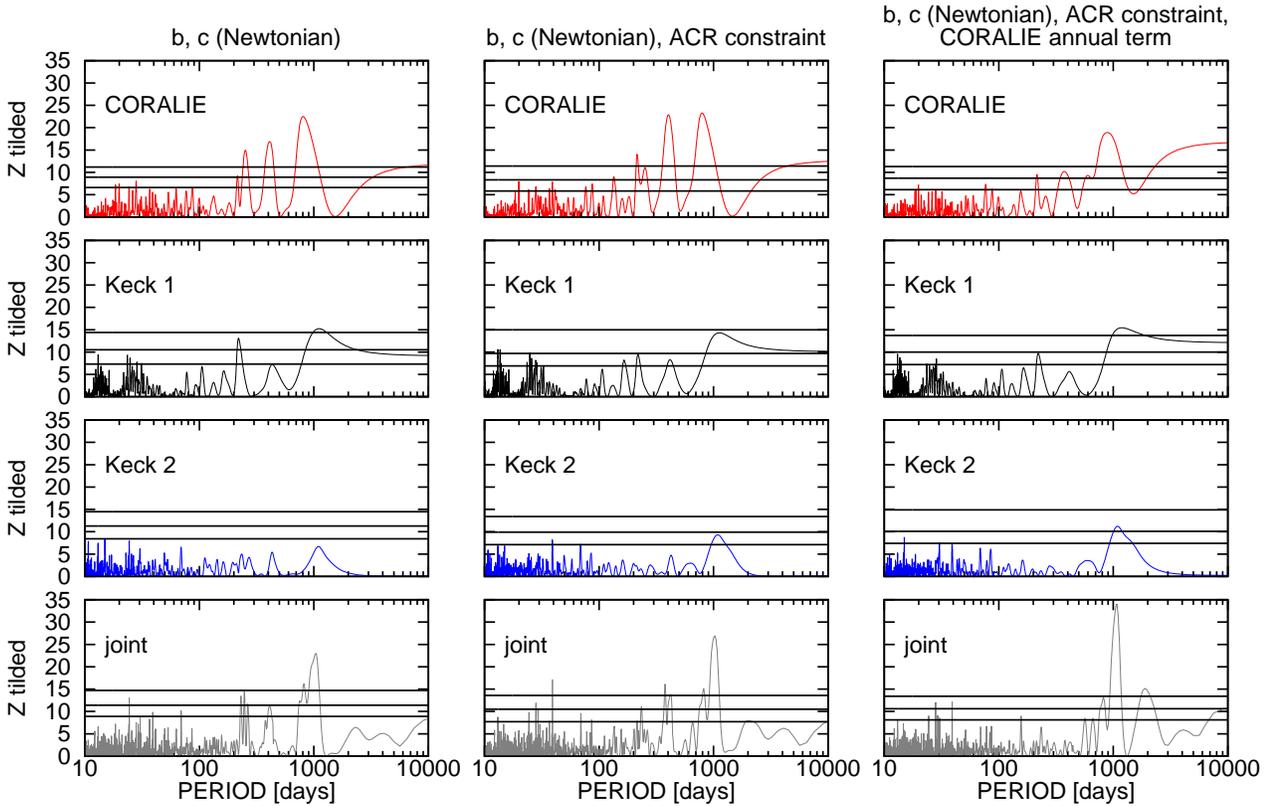}
\caption{Residual likelihood-ratio periodograms of the HD82943 RV data for different
base models, which are marked in the column titles. The periodograms correspond to the
entire time series and to specific sub-data-sets (labelled in the graphs). The latter ones
were constructed assuming an RV model with a probe periodicity assigned to only a
particular RV data-set (labelled in each individual panel). The minimum period was increased
to $10$~d from the traditional limit of $1$~d, because the time values of the CORALIE data
were relatively inaccurate, as these data were scanned from a figure.}
\label{fig_pow}
\end{figure*}

To clarify the situation, let us consider some residual periodograms calculated with
respect to different base models. They are plotted in Fig.~\ref{fig_pow}. Here we plot
periodograms related to each of the individual data sets as well as to the joint time
series. The data-set-separated periodograms were constructed according to the method
described in \citet{Baluev11}. Namely, we assume that the probe periodic variation belongs
to only a single RV data set, while the base RV model (e.g. the ones of Table~\ref{tab_bc}
or Table~\ref{tab_bc_ACR}) still belongs to each of them. This approach allows to easily
detect various inconsistencies between different data sets, still using the full
statistical power of the entire time series to fit the base model. See also
\citep{Baluev13c} and references therein for a more unified description of the ``residual
periodograms'' that we use here. Also, we note that when computing these periodograms the
common orbital inclination was allowed to float to absorb as much of the residual
variation as possible. The horizontal lines in the graphs of Fig.~\ref{fig_pow} show the
simulated statistical levels of $1$-sigma, $2$-sigma, and $3$-sigma significance, which we
discuss in more details in Section~\ref{sec_valid} below.

\begin{figure}
\includegraphics[width=84mm]{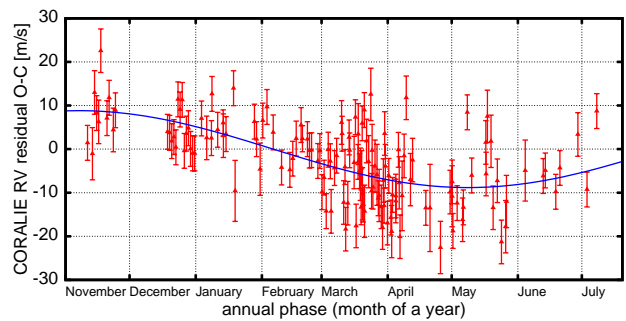}
\caption{The best fitting CORALIE annual variation phased to its period. The RV model here
is the three-planet Newtonian one with a fittable inclination. The RV contribution of the
planets was subtracted from the residuals before plotting the graph, but the contribution
from the annual variation was preserved.}
\label{fig_rsdY}
\end{figure}

Looking at Fig.~\ref{fig_pow}, we can draw the following conclusions:
\begin{enumerate}
\item The two-planet RV model is definitely unable to explain the full RV variation in the
data. There are one or even more periodic or quasi-periodic variations in the residuals.

\item One of the main differences between the CORALIE and Keck data sets is an RV
variation at the period of $\sim 400$~d. This is the only large peak that is significantly
pumped up when we set up the ACR constraint (i.e. when we analyse RV residuals
corresponding to a more realistic orbital model). Therefore, it is the likely source of
the inconsistency.

\item Since the $\sim 400$~d period is close to the second planet's period, it might be
tempting to interpret it as some artifact of an incomplete reduction of the RV variation
due to the planets \emph{c} and \emph{b}. This is however unlikely, because then it should
exist in all three RV data sets. We interpret it as an annual variation caused by
instrumental or data reduction errors related only to the CORALIE data. The periodogram
peak is wide enough to be consistent with the period of $365$~d. It is already known that
annual errors frequently occur in the old ELODIE RV data \citep{Baluev08b,Baluev08c}; the
same may be true for CORALIE. We plot this CORALIE annual term in Fig.~\ref{fig_rsdY}

\item Another periodogram peak persists with a period of $\sim 1100$~d. On contrary to the
$400$~d peak, this one is present (and is statistically significant) in all three RV
data-sets, regardless of whether we consider them jointly or separately. After removal of
the CORALIE annual variation, the $\sim 1100$~d one even becomes more obvious. This
convinces us that some RV variation at the period of $\sim 1100$~d does exists and it
belongs to the star rather than to a specific instrument.
\end{enumerate}

The RV variation near the $1100$~days period was already suspected in previous studies
\citep{Beauge08}. Its most intriguing explanation, that we analyse further in this work,
is the possible existence of a third planet in the system. This hypothetical third planet
would be very interesting because it appears close to the 5:2 MMR with the planet
\emph{b}. Thus the whole system would lie close to the three-planet 1:2:5 resonance. This
is different from the 1:2:4 (Laplace) resonance suggested by \citet{Beauge08}. The updated
RV data no longer support the Laplace resonance. The  Tables~\ref{tab_bcdY}
and~\ref{tab_bcdY_ACR} contain the parameters of the three-planet fits that were obtained
with and without the ACR constraint. Both fits correspond to an aligned apsidal corotation
between the planets \emph{c} and \emph{b}. We interpret this as a sign of a considerably
more ``healthy'' RV model. Although the non-ACR configuration of Table~\ref{tab_bcdY}
still appears formally unstable, it is now very easy to slightly adjust its parameters to
make it stable. In fact, now the ACR and non-ACR fits are statistically consistent with
each other: we obtain only $1.5$-sigma separation from the likelihood-ratio test.

\begin{table}
\caption{Best fitting parameters of the HD82943 planetary system: three-planet Newtonian
edge-on model with the CORALIE annual term}
\label{tab_bcdY}
\begin{tabular}{llll}
\hline
\multicolumn{4}{c}{planetary orbital parameters and masses} \\
                       & planet c         & planet b       & planet d       \\
$P$~[day]              & $220.080(70)$    & $439.70(48)$   & $1078(13)$     \\
$K$~[m/s]              & $58.5(2.3)$      & $39.31(55)$    & $5.30(57)$     \\
$e$                    & $0.410(16)$      & $0.053(63)$    & $0 \rm (fixed)$\\
$\omega$~[$^\circ$]    & $117.1(1.2)$     & $123.5(9.7)$   & $-$            \\
$\lambda$~[$^\circ$]   & $307.3(1.1)$     & $215.00(97)$   & $296.0(6.0)$   \\
$M$~[$M_{\rm Jup}$]    & $1.721(78)$      & $1.593(21)$    & $0.290(31)$    \\
$a$~[AU]               & $0.74340(16)$    & $1.17978(86)$  & $2.145(17)$    \\
$i$~[$^\circ$]         &\multicolumn{3}{c}{$90 \rm (fixed)$} \\
\hline
\multicolumn{4}{c}{parameters of the data sets} \\
                           & CORALIE        & Keck 1          & Keck 2      \\
$c$~[m/s]                  & $8146.1(1.1)$  & $-4.8(1.1)$     & $-6.86(51)$ \\
$A_{\rm sys}$~[m/s]        & $8.5(1.3)$     &                 &             \\
$\tau_{\rm sys}$~[day]     & $184.8(9.9)$   &                 &             \\
$\sigma_{\rm jitter}$~[m/s]& $5.63(55)$     & $4.90(81)$      & $2.61(34)$  \\
r.m.s.~[m/s]               & $7.13$         & $4.87$          & $2.76$      \\
\hline
\multicolumn{4}{c}{general characteristics of the fit} \\
$\tilde l$~[m/s]           & \multicolumn{3}{c}{$5.86$} \\
$d$                        & \multicolumn{3}{c}{$21$}   \\
\hline
\end{tabular}\\
See notes of Table~\ref{tab_bc}. The additional parameters $A_{\rm sys}$ and $\tau_{\rm
sys}$ represent the semiamplitude and the maximum epoch of the sinusoidal CORALIE annual
variation.
\end{table}

\begin{table}
\caption{Best fitting parameters of the HD82943 planetary system: three-planet Newtonian
edge-on ACR(\emph{c},\emph{b}) model with the CORALIE annual term}
\label{tab_bcdY_ACR}
\begin{tabular}{llll}
\hline
\multicolumn{4}{c}{planetary orbital parameters and masses} \\
                       & planet c         & planet b       & planet d       \\
$P$~[day]              & $220.062(33)$    & $439.586(74)$  & $1072(13)$     \\
$K$~[m/s]              & $55.22(55)$      & $39.86(56)$    & $5.39(57)$     \\
$e$                    & $0.4289(92)$     & $0.1476(50)$   & $0 \rm (fixed)$\\
$\omega$~[$^\circ$]    & $118.0(1.1)$     & $118.0(1.1)$   & $-$            \\
$\lambda$~[$^\circ$]   & $309.10(49)$     & $213.56(71)$   & $298.4(5.8)$   \\
$M$~[$M_{\rm Jup}$]    & $1.607(18)$      & $1.600(22)$    & $0.294(31)$    \\
$a$~[AU]               & $0.743340(73)$   & $1.17955(13)$  & $2.137(17)$    \\
$i$~[$^\circ$]         &\multicolumn{3}{c}{$90 \rm (fixed)$} \\
\hline
\multicolumn{4}{c}{parameters of the data sets} \\
                           & CORALIE        & Keck 1          & Keck 2      \\
$c$~[m/s]                  & $8146.1(1.1)$  & $-4.8(1.0)$     & $-6.89(57)$ \\
$A_{\rm sys}$~[m/s]        & $8.9(1.2)$     &                 &             \\
$\tau_{\rm sys}$~[day]     & $182.4(9.3)$   &                 &             \\
$\sigma_{\rm jitter}$~[m/s]& $5.68(55)$     & $4.16(71)$      & $2.99(38)$  \\
r.m.s.~[m/s]               & $7.21$         & $4.24$          & $3.10$      \\
\hline
\multicolumn{4}{c}{general characteristics of the fit} \\
$\tilde l$~[m/s]           & \multicolumn{3}{c}{$5.91$} \\
$d$                        & \multicolumn{3}{c}{$21-4=17$}   \\
\hline
\end{tabular}\\
See notes of Table~\ref{tab_bc} and Table~\ref{tab_bc_ACR}.
\end{table}

Here we should note that the ACR configuration of the two inner planets is perturbed by
the third planet. Formally, this perturbation should slightly shift the parameters of the
ACR configuration from its purely two-planet position. However, we do not take this ACR
shift into account in Table~\ref{tab_bcdY_ACR}, because it is expected to be pretty small.
A rough estimation yields that the gravitational force from the planet \emph{d} is smaller
than $1/10$ of the one existing between the planets \emph{c} and \emph{b}. Therefore, the
ACR constraint of Table~\ref{tab_bcdY_ACR} was based only on the Hamiltonian of the
subsystem of the two inner planets, as if the third planet had no influence. From the
other side, when we obtained the both fits of Tables~\ref{tab_bcdY}
and~\ref{tab_bcdY_ACR}, the gravitational perturbation from the third planet was still
taken into account in full to compute the fitted RV model. In either case, dynamical
simulations show that the fit of Table~\ref{tab_bcdY_ACR} is very close to the desired ACR
state, so this fit is the one that we expected to obtain.

\begin{figure*}
\includegraphics[width=0.95\linewidth]{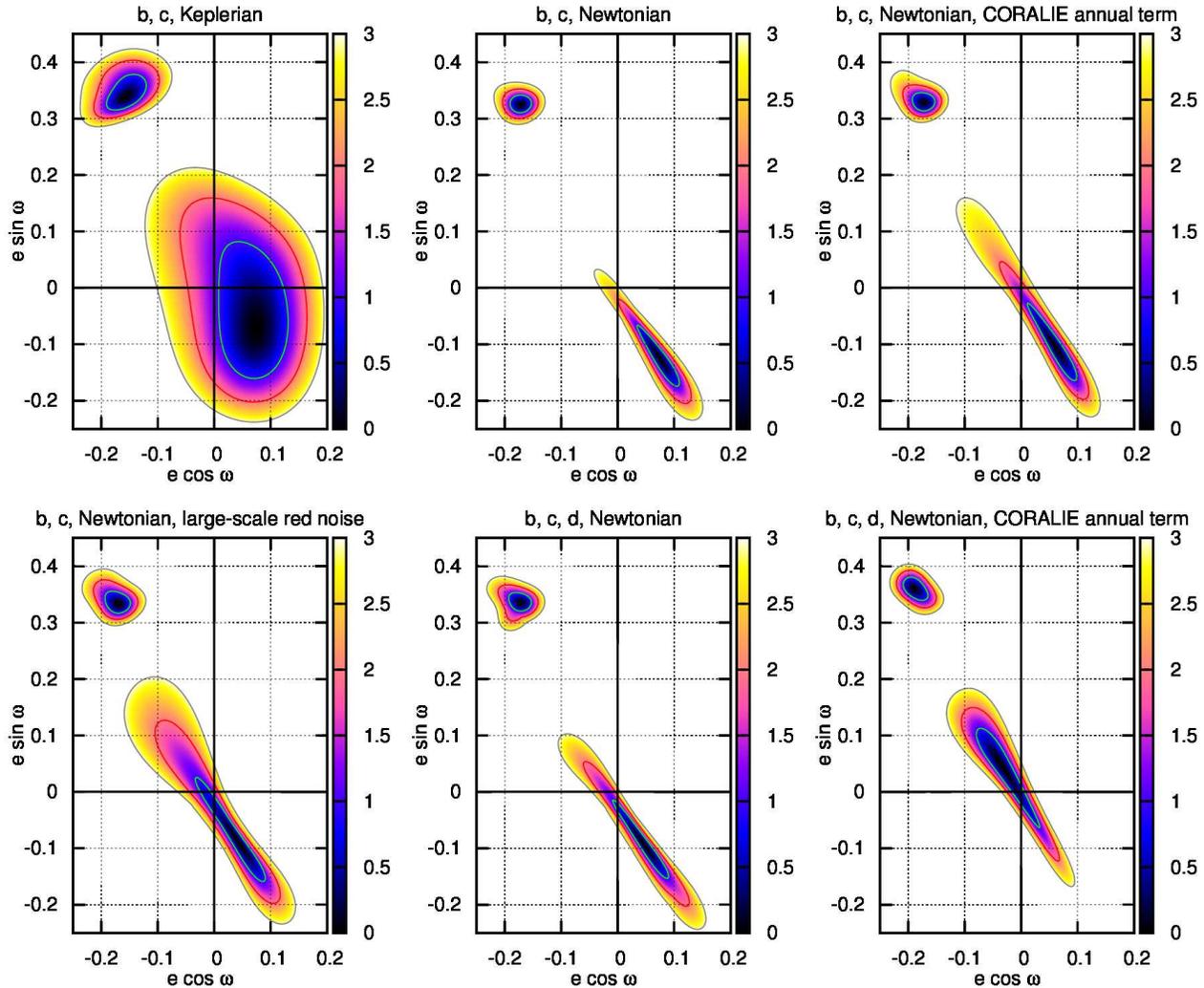}
\caption{Confidence regions for the parameters $(e\cos\omega,e\sin\omega)$ of the HD82943
planets \emph{c} and \emph{b} for various RV data models. In each panel, the smaller spot
in the left-top part of graph is for the planet \emph{c}, and the large spot is for the
planet \emph{b}. The isolines correspond to the asymptotic $1$-sigma, $2$-sigma, and
$3$-sigma significance levels. The type of the orbital model is marked in the titles above
each plot. For the Newtonian fits we always used a coplanar orbital model with a fittable
common inclination. For three-planet models the planet \emph{d} was included in the
$N$-body integration, but its eccentricity was always fixed at zero. See text for further
details and discussion.}
\label{fig_we}
\end{figure*}

In view of the matters discussed above, it is interesting to track how the CORALIE annual
variation and the RV contribution from the putative $\sim 1100$~days planet could distort
the best fitting parameters of the main planets \emph{c} and \emph{b}. To do this, we
consider two-dimensional confidence regions for the parameters $(e\cos\omega,
e\sin\omega)$ that are plotted in Fig.~\ref{fig_we}. We give these plots for different
models of the RV data. Here is what we would like to highlight:
\begin{enumerate}
\item The transition from the Keplerian to Newtonian RV model has a dramatic shrinking
effect on the uncertainty region of $(e_b,\omega_b)$. However, this shrinking is severely
anisotropic, and the best fitting values themselves remain rather immutable.

\item The only way to naturally obtain a stable two-planet configuration, without the use
of any ``brute force'' like an imposed ACR constraint, is to take into account the
residual long-term RV variations discussed above.

\item The stable two-planet configuration can be approached to in many ways: by adding a
fittable annual variation to the CORALIE RV model, by adding the $\sim 1100$~days periodic
term, or by dealing with the red-noise model using the method of \citet{Baluev13a}. Either
of these actions shifts the best fitting orbital configuration in the direction of the
stable aligned ACR. However, the largest effect is obtained when the CORALIE annual term
and the third planet are both included. In this case, the best fit itself migrates to the
domain of aligned apses, and although this best fit is still unstable, it becomes easy to
find stable configurations within its statistical uncertainties.
\end{enumerate}

\section{Statistical validity of the results}
\label{sec_valid}
So far we mainly relied on analytical likelihood-ratio tests with its asymptotic
chi-square distribution, and on the maximum-likelihood point estimations that are valid
when the number of observations is sufficiently large. These methods are often criticised
by authors who propagate the use of the Bayesian statistical methods instead. However, we
believe that the weaknesses of Asymptotic Maximum-Likelihood Theory~-- AMLET~-- are often
excessively exacerbated, as well as the advantages of the Bayesianism which is suggested
as a replace. This has been already demonstrated in \citet{Baluev13a} for the case of the
GJ581 planetary system, in which a rather complicated multi-planet model with correlated
noise was employed. Here we aim to show that the case of HD82943 is similar in this
concern, at least when all publicly available RV data are taken in the analysis.

We rely here on the Monte Carlo simulations assuming the Gaussian model of the RV noise.
We do not believe that employing the bootstrap simulation, like \citet{Tan13}, would be
more reliable. From Fig.~\ref{fig_rsd} it is clear that the residuals to the best fitting
models are anyway corrupted by some systematic variations, so it is perhaps useless to
expect that shuffling of these corrupted residuals would provide a better model of the
real RV noise. Besides, from \citet{Baluev13a} we know that the bootstrap method does not
correctly handle the uncertainty of noise parameters like the jitter.

So, let us verify that we indeed can safely treat various likelihood-ratio tests obeying
to the asymptotic chi-square distribution. To do this we must adopt some null hypothesis
$\mathcal H$ (in the form of the functional model) and the alternative hypothesis
$\mathcal K$, such that $\mathcal H$ is a subspace of $\mathcal K$ of lesser dimension.
The simplest case, for example, is when we assume that $\mathcal H$ consists of a single
point, while $\mathcal K$ spans the entire parametric space (for some given parametric
model of the RV data). After choosing the models $\mathcal H$ and $\mathcal K$ we can run
the Monte Carlo simulation (the PlanetPack algorithm described in Sect.~10.1 of
\citealt{Baluev13c}) to reconstruct the distribution of the associated non-linear
likelihood-ratio statistic and to compare it with the relevant chi-square distribution.

\begin{figure*}
\includegraphics[width=\textwidth]{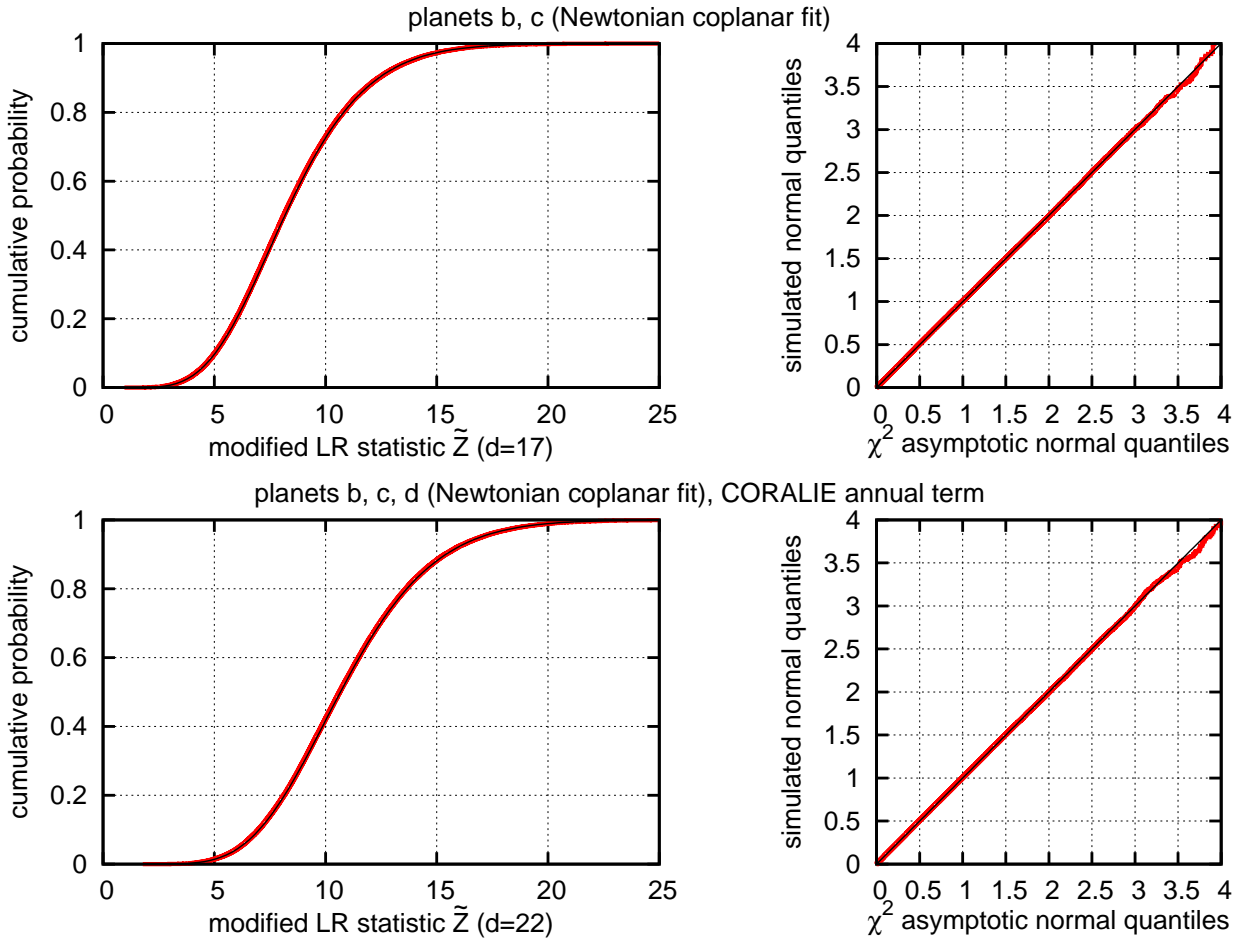}
\caption{The distributions of the test statistic $\tilde Z$ from \citep{Baluev08b}:
comparing the asymptotic $\chi^2$ one (thick red curves) and the simulated one (thin black
curves) in each graph. Left graphs show the cumulative distribution functions (with the
number of $\chi^2$ degrees of freedom, $d$, labelled in the abscissa). Right graphs
compare the relevant normal quantiles (the $n\sigma$ significance levels). Here, the null
model $\mathcal H$ is a single point related to some adopted ``true'' parameters (actually
borrowed from the best fit of the \emph{real} RV data), and the alternative model
$\mathcal K$ involves a full fit of the simulated data.}
\label{fig_mc_main}
\end{figure*}

In Fig.~\ref{fig_mc_main} we compare the simulated and analytic likelihood-ratio
distributions for the case when $\mathcal H$ is a single point (which is treated as the
true vector of the parameters), while $\mathcal K$ corresponds to a two-planet or a
three-planet model (with a fittable common inclination in each case). We can see that the
agreement is very accurate, as if these models were strictly linear.

\begin{figure*}
\includegraphics[width=\textwidth]{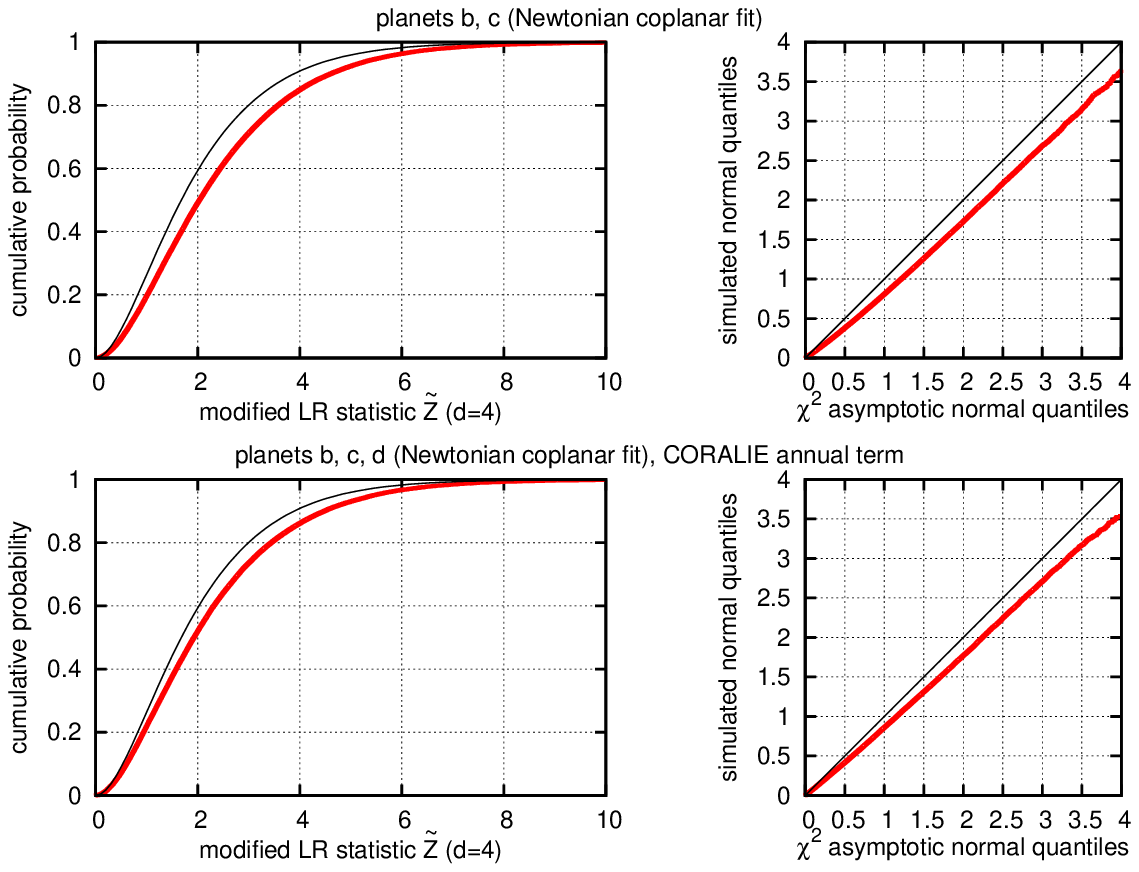}
\caption{Same as Fig.~\ref{fig_mc_main}, but for testing an ACR best fit against the
corresponding unconstrained best fit.}
\label{fig_mc_ACR}
\end{figure*}

Proceeding further, we verify the applicability of the asymptotic likelihood-ratio test to
the task of comparing the ACR with a general non-ACR configuration. The simulation results
for this test are shown in Fig.~\ref{fig_mc_ACR}. In this case the alternative $\mathcal
K$ represents the entire parametric space (for the same two models as above), while the
null hypothesis $\mathcal H$ is a restriction of $\mathcal K$ to the ACR models. Now we
can see some deviation between the simulated and asymptotic distribution function, but
this deviation is rather marginal. In particular, the $\sim 1.5\sigma$ statistical
difference between the fits of Tables~\ref{tab_bcdY} and~\ref{tab_bcdY_ACR} is now
corrected to $\sim 1.3\sigma$, which is similarly insignificant. The difference of $\sim
4.2\sigma$ between the fits of Tables~\ref{tab_bc} and~\ref{tab_bc_ACR} is slightly
reduced to $\sim 3.8\sigma$, which is still very large. Therefore, these correction are
rather cosmetic and do not trigger any qualitative changes.

\begin{figure*}
\includegraphics[width=\textwidth]{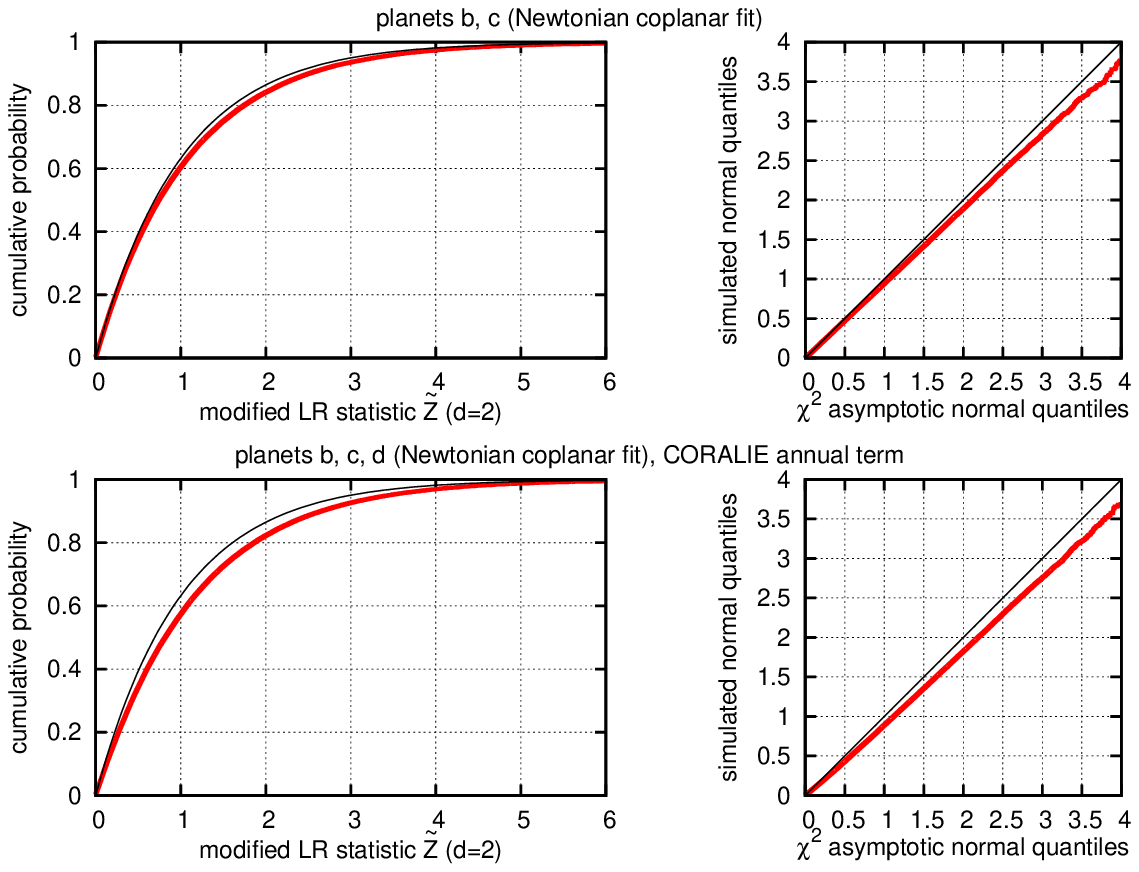}
\caption{Same as Fig.~\ref{fig_mc_main}, but for testing a best-fit model with fixed
$(e_b,\omega_b)$ against the corresponding unconstrained fit. These results can be used
for a more accurate calibrations of the confidence contours in Fig.~\ref{fig_we}.}
\label{fig_mc_we2}
\end{figure*}

Finally, we verify the calibration of the confidence contours that we plotted in
Fig.~\ref{fig_we}. These confidence regions represent the level contours of the likelihood
function, and they can also be treated by means of the likelihood-ratio test
\citep{Baluev13c}. In this case the alternative hypothesis again fills the entire
parametric space, while the null hypothesis represents its restriction to some fixed
values of $\omega$ and $e$ (with unrestricted other parameters). In this case the
deviation between the distribution functions is even smaller than for the previous ``ACR
vs. non-ACR'' comparison. This means that the relevant corrections to Fig.~\ref{fig_we}
would be rather unremarkable, so these confidence regions are statistically safe and
reliable.

The case of the periodogram distributions did not appear that nice, however. The
periodogram analog of the asymptotic chi-square likelihood-ratio distribution is the
following approximation given in \citep{Baluev08a}:
\begin{equation}
\FAP(z) \lesssim M(z) \approx W e^{-z} \sqrt z,
\label{prdg_fap}
\end{equation}
where $\FAP$ is the false alarm probability to estimate, $z$ is the observed periodogram
maximum, and $W$ is proportional to the settled frequency range. Formally, this formula is
strictly valid only for linear models (with a single allowed non-linear frequency
parameter), but for the non-linear periodograms it should be still valid in an asymptotic
sense for $N\to\infty$ \citep{Baluev08b}.

\begin{figure}
\includegraphics[width=84mm]{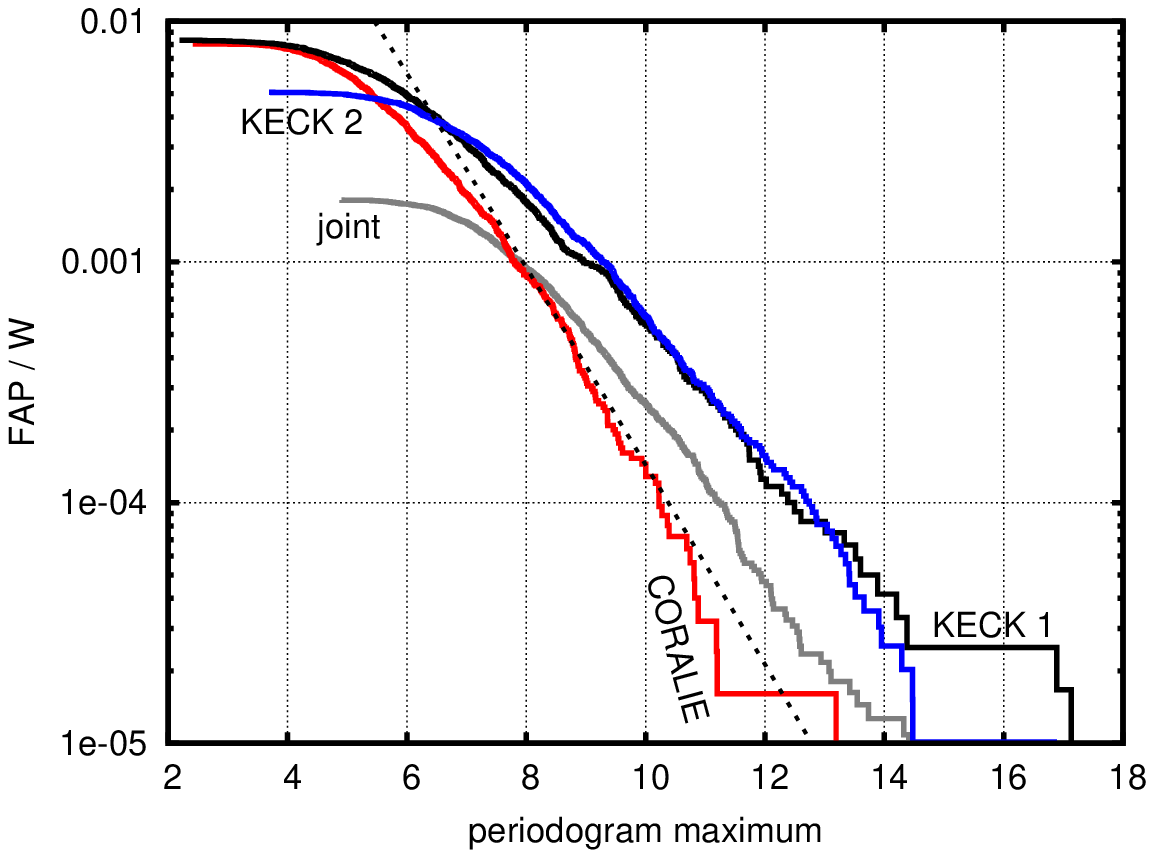}
\caption{Simulated distributions of the periodogram maxima for periodograms in the left
column of Fig.~\ref{fig_pow}. The ordianate is equal to the simulated $\FAP$ divided by
$W=f_{\rm max} T_{\rm eff}$, where $f_{\rm max}$ was always $0.1$~d$^{-1}$, while $T_{\rm
eff}$ was different: $5534$~d for the joint data, $1245$~d for CORALIE, $1200$~d for
Keck-1, and $1974$~d for Keck-2. The broken line is the graph of the function
$e^{-z}\sqrt{z}$, a common asymptotic approximation based on~(\ref{prdg_fap}). Number of
Monte Carlo trials was equal to $1000$.}
\label{fig_mc_pow}
\end{figure}

However, the large factor $W$ in~(\ref{prdg_fap}) scales up any non-linearity effects in
the $\FAP$ to levels much larger than e.g. the small deviations seen in
Figs.~\ref{fig_mc_main}, \ref{fig_mc_ACR}, and~\ref{fig_mc_we2} above. Monte Carlo
simulations have shown that the approximation~(\ref{prdg_fap}) works well only for the
CORALIE periodogram, while for the Keck periodograms the simulated $\FAP$ is much larger
than~(\ref{prdg_fap}) predicts. This was actually expected, since the number of the Keck
data is still rather small, and they are split in two even smaller independent subsets.
Therefore, we decided to calibrate the periodogram significance levels of
Fig.~\ref{fig_pow} with the simulated values of the $\FAP$ rather than with the analytic
approximation~(\ref{prdg_fap}). These simulations were done for the frequency range from
$0$ to $0.1$~d$^{-1}$, which is the same that was used for the periodograms themselves.
The simulated $\FAP$ curves are shown in Fig.~\ref{fig_mc_pow}.

\section{Reality of the third planet}
\label{sec_third}
\begin{figure}
\includegraphics[width=84mm]{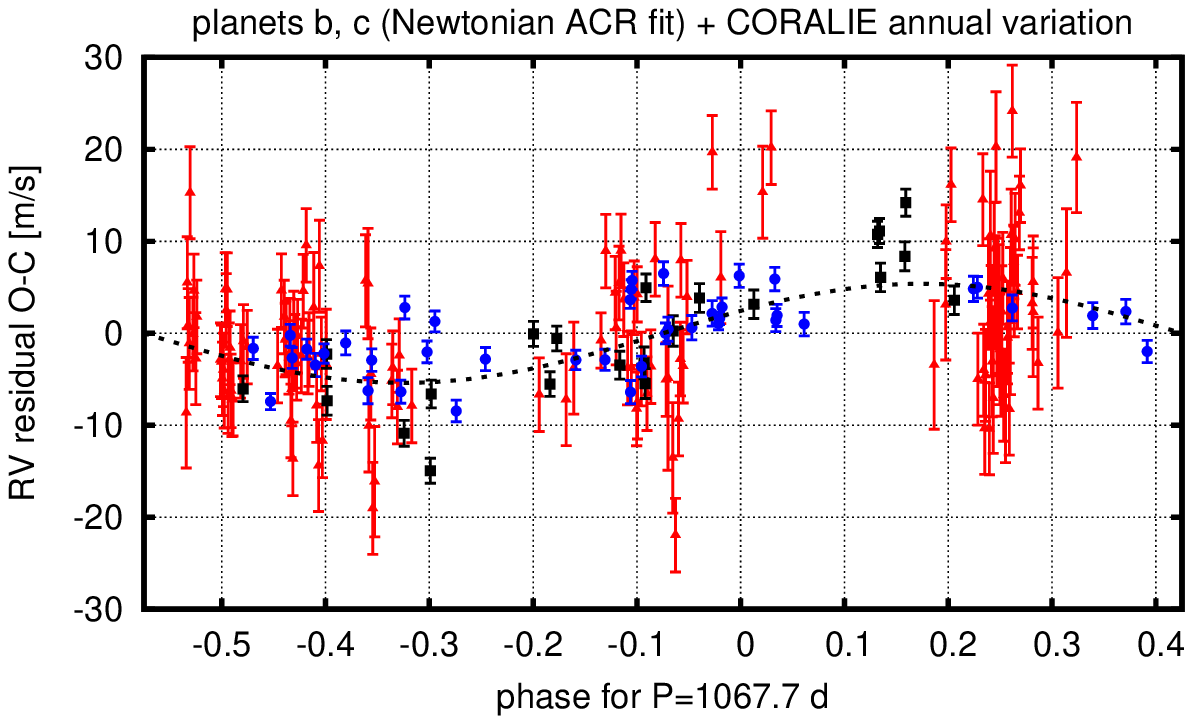}
\caption{The two-planet RV residuals of HD82943 phased to the \emph{apparent} RV period of
the third planet (which is by a few days smaller than the best fitting \emph{osculating}
orbital period). The reference fit corresponds to the ACR two-planet Newtonian model with
a fittable inclination. The data points have the same shapes as in Fig.~\ref{fig_rsd}.}
\label{fig_rsd_d}
\end{figure}

A strong evidence in favour of the $1100$~d variation comes from its detectability in all
three RV data sets that we have analysed: the CORALIE data, the Keck data before the
upgrade, and the Keck data after the upgrade. Although the RV data coverage is not
entirely perfect (the middle $1100$~d cycle was poorly covered), the phase of this
sinusoid is more or less smoothly transferred from one data array to another (see
Fig.~\ref{fig_rsd} and~\ref{fig_rsd_d}). Besides, it appears difficult to naturally obtain
a stable two-planet configuration from the combined time series, unless the $1100$-day
variation is taken into account. This argumentation suggests that the mentioned variation
does really exist and is likely caused by the star rather than by some systematic
instrumental drifts or data reduction errors.

Non-planetary interpretations of the $\sim 1100$-day RV variability are still possible.
This variation could be caused by some long-term astrophysical activity phenomenon
evolving on the star. In particular, the long-term stellar magnetic activity is known to
generate excessive noise in the low-frequency range \citep{Dumusque12}. In fact, it looks
rather suspicious that this variation seems to fade over years: it was strong in the time
of CORALIE observations, while in the latest Keck data it is more difficult to detect.
However, this seems to be an apparent effect due to a more dense CORALIE data coverage and
their larger number, since in the phased residuals (Fig.~\ref{fig_rsd_d}) we do not see
any clear systematic differences between different data sets.

We tried to verify the long-term noise hypothesis using the red-noise analysis technique
described in \citet{Baluev13a}. Our result is that the red-noise model can easily absorb
the $\sim 1100$-day variation, suppressing it well below the characteristic noise levels.
The characteristic correlation timescale that the red-noise analysis algorithm reported
was about a few hundred of days. Therefore, the current data are unable to distinguish the
planetary and non-planetary interpretations. This ambiguity may be resolved analysing the
correlation between the Doppler RV measurements and some spectral activity indicator, as
\citet{Dumusque12} have done for $\alpha$~Cen. If the same $1100$~d period is found in the
star's activity measure than we should probably retract the hypothesis of the third
planet. The necessary spectral data are not publicly available, however. In any case, we
still must take the relevant RV variation into account to have a more robust two-planet
fit.

\begin{figure}
\includegraphics[width=84mm]{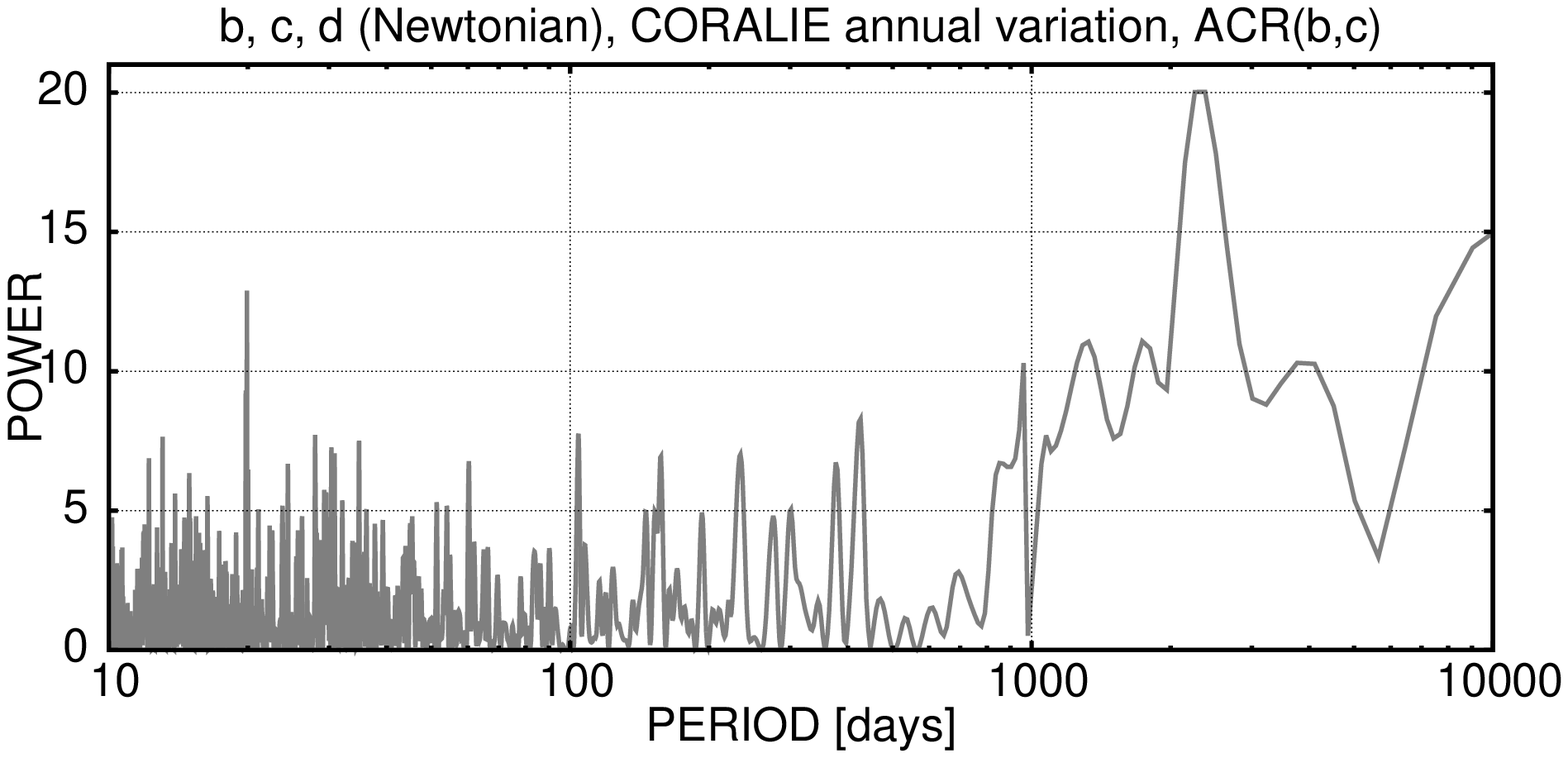}
\caption{Residual periodogram calculated for the three-planet ACR model of HD~82943, based on
the joint RV data. The system inclination to the sky plane is treated as a fittable
parameter. Analogous non-ACR base model generates a similar periodogram with a slightly
reduced peak at $2300$~d.}
\label{fig_powrsd}
\end{figure}

In fact, both interpretations can work together: the $1100$-day variation may be induced
by the third planet indeed, and it might be also contaminated by the long-term
astrophysical noise. To verify this possibility, we computed the residual periodogram with
all three planets included in the base model (see Fig.~\ref{fig_powrsd}). We can see that
some residual power at long periods still remains, and it might be even statistically
significant. Hwever it looks like some large-scale noise rather than a single clearly
isolated period. Besides, it is remarkably smaller than the planet~\emph{d} peak that
remained in the residuals of the analogous two-planet model (right-bottom panel of
Fig.~\ref{fig_pow}). We prefer to interpret the residual power in Fig.~\ref{fig_powrsd} as
some astrophysical noise or remaining systematic instrumental errors. We definitely need
more RV data to investigate this remaining residual variation more reliably.

\begin{figure}
\includegraphics[width=84mm]{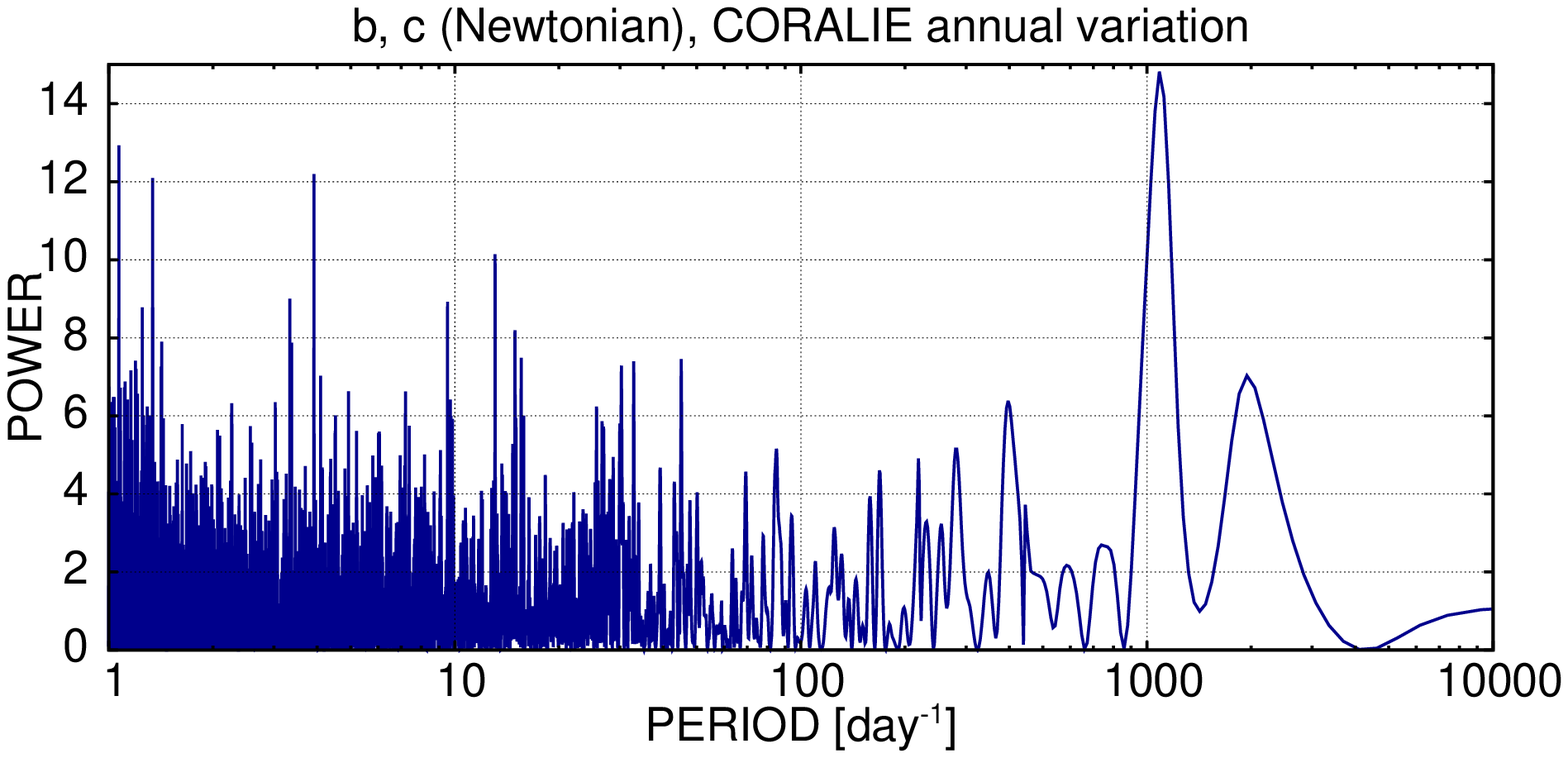}
\caption{Residual periodogram calculated for the two-planet non-ACR model of
HD~82943, based on the standalone Keck RV data (without CORALIE). The system inclination
to the sky plane, $i$, is treated as a fittable parameter. The models with $i$ fixed at
$90^\circ$ and at $20^\circ$ generate periodograms with a slightly higher peak at $\sim
1100$~d.}
\label{fig_powKECK}
\end{figure}

\begin{figure}
\includegraphics[width=84mm]{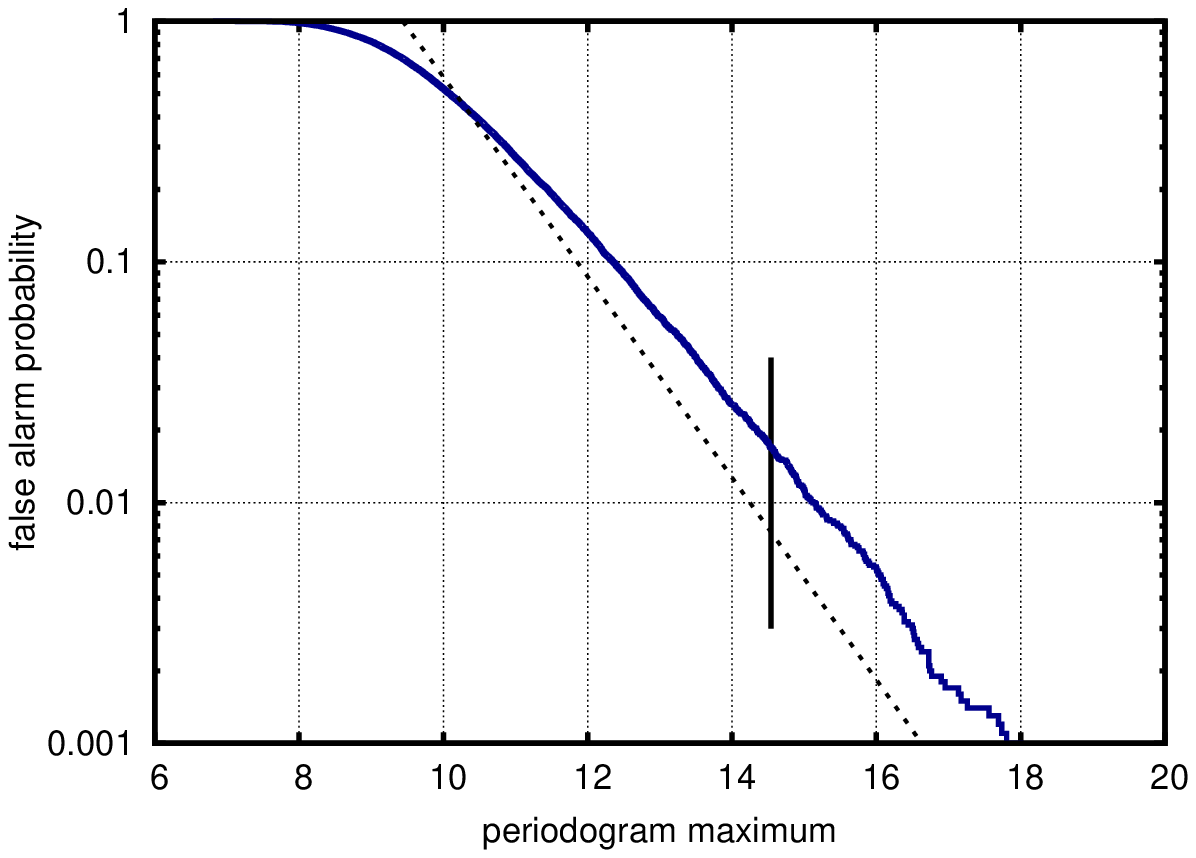}
\caption{The solid (blue) line shows the simulated false alarm probability for the periodogram
in Fig.~\ref{fig_powKECK}. The broken line shows asymptotic
approximation~(\ref{prdg_fap}). The vertical line segment labels the abscissa position
corresponding to the height of the actual periodogram peak at $P\sim 1100$~d. Number of
Monte Carlo trials was equal to $10000$.}
\label{fig_mc_powK}
\end{figure}

\citet{Tan13} also noted a peak at $\sim 1100$~d in their Keck periodograms. However,
their statistical analysis yielded only a marginal significance for this variation:
$\FAP=0.033$ for an edge-on fit and $\FAP=0.085$ for a fit with $i=20^\circ$. Our work
yields a remarkably more credible detection. To carry out a more direct comparison with
\citet{Tan13}, in addition to the periodograms of Fig.~\ref{fig_pow} we have also computed
the periodogram of the Keck data that were taken entirely alone (without CORALIE) and
without imposing of any ACR constraint. The analytic formula~(\ref{prdg_fap}) implied
$\FAP \sim 10^{-3}$ for the edge-on model, $\FAP \sim 4 \times 10^{-3}$ for the
$i=20^\circ$ model, and $\FAP \sim 6 \times 10^{-3}$ for the model with a floating $i$
(which still appeared close to $20^\circ$). Note that these $\FAP$ values are for the
frequency range from $0$ to $1$~d$^{-1}$, which is $10$ times wider than that of
Fig.~\ref{fig_pow}. However, we have explained above that the analytic
formula~(\ref{prdg_fap}) underestimates the $\FAP$ for Keck periodograms. Therefore, we
have selected the worst case of three~--- the model with free $i$~--- and run the Monte
Carlo simulation to assess the relevant $\FAP$ more reliably. We obtain the estimation of
$\FAP\sim 0.016$. The relevant periodogram is shown in Fig.~\ref{fig_powKECK}, and the
graph of the associated simulated $\FAP$ is shown in Fig.~\ref{fig_mc_powK}.

This suggests remarkably more credible detection of the planet~\emph{d} than what follows
from \citet{Tan13} results: e.g. the $\FAP$ is now reduced by the factor of $5$ or more.
Although our Keck periodogram in Fig.~\ref{fig_powKECK} should be even more pessimistic
than the both periodograms shown by \citet{Tan13} in their Fig.~13, our $\sim 1100$~d peak
is somewhat higher (considering all the cases relatively to their apparent noise levels).
Such difference was caused, as we believe, by the following main factors. First, we used
the more efficient ``residual periodogram'' (also known as ``recursive periodogram'')
instead of the ``periodogram of the residuals''. The so-called residual periodogram is
based on a full multi-planet fit per \emph{each} computed power value, rather than on a
single fit of the base model. Thus we deal with more adequate and accurate fits, which
improve the periodogram detection power by pushing the real peaks up relatively to the
noisy ones \citep{Anglada-Escude12,Baluev13c}. Secondly, we used a more adequate RV noise
model which involves an adaptive jitter fitting, and also allows for a more reliable
relative weighting of different data sets
\citep{Baluev08b}.

Our simulation of the Keck-only detection $\FAP$ of $1.7$ per cent (or $2.4\sigma$) for
the planet~\emph{d} is still slightly worse than $1$ per cent (or $2.6\sigma$), which
\citet{Tan13} acknowledge as a trustable exoplanetary detection threshold. However, we
re-emphasize that this long-term variation is also supported by the public CORALIE data
(even after reduction of their annual variation), and the cumulative significance is much
better than even the $3\sigma$ level (see Fig.~\ref{fig_pow}).

\section{Uncertainty of the inclination}
\label{sec_inc}
During the preparation of this manuscript, a new observational work has appeared
\citep{Kennedy13}, where the authors consider the debris disk of HD82943 (which was
originally discovered by \citet{Beihmann05}) and report a rather accurate measurement of
its inclination to the sky plane: $27^\circ \pm 4^\circ$. They find this value in a good
agreement with the planetary system inclination $i=20^\circ\pm 4^\circ$, as reported by
\citet{Tan13} based on the Keck RV fits. Contrary to \citet{Tan13}, the orbital
inclination estimations of our work are very uncertain, with the lower typical limit on
$i$ about $\sim 20^\circ$ and no upper limit (i.e., consistent with an edge-on
orientation, $i=90^\circ$). The typical nominal value is $i \sim 40^\circ$. These results
are in fact also consistent with \citet{Kennedy13}, at least we do not find any detectable
disagreement. We only have to be a bit more sceptical concerning the conclusion that the
disk-planets alignment is confirmed indeed; our RV data analysis just does not allow to
verify this guess.

We believe that the existence of an observable outer debris disk in the system tells us
indirectly that the space beyond the two robustly detectable planets is unlikely empty:
the third planet or even more additional planets may be present.

\begin{table}
\caption{Best fitting parameters of the HD82943 planetary system: three-planet Newtonian
inclined ACR(\emph{c},\emph{b}) model with the CORALIE annual term}
\label{tab_bcdY_inc_ACR}
\begin{tabular}{llll}
\hline
\multicolumn{4}{c}{planetary orbital parameters and masses} \\
                       & planet c         & planet b       & planet d       \\
$P$~[day]              & $220.158(33)$    & $439.16(10)$   & $1075(13)$     \\
$K$~[m/s]              & $55.41(54)$      & $39.92(55)$    & $5.42(57)$     \\
$e$                    & $0.4257(93)$     & $0.1460(50)$   & $0 \rm (fixed)$\\
$\omega$~[$^\circ$]    & $118.0(1.1)$     & $118.0(1.1)$   & $-$            \\
$\lambda$~[$^\circ$]   & $309.55(50)$     & $213.77(72)$   & $293.5(5.7)$   \\
$M$~[$M_{\rm Jup}$]    & $3.559(40)$      & $3.529(48)$    & $0.653(70)$    \\
$a$~[AU]               & $0.743963(73)$   & $1.18006(19)$  & $2.144(17)$    \\
$i$~[$^\circ$]         & \multicolumn{3}{c}{$27 \rm (fixed)$} \\
\hline
\multicolumn{4}{c}{parameters of the data sets} \\
                           & CORALIE        & Keck 1          & Keck 2      \\
$c$~[m/s]                  & $8146.3(1.2)$  & $-4.79(91)$     & $-6.73(55)$ \\
$A_{\rm sys}$~[m/s]        & $8.7(1.2)$     &                 &             \\
$\tau_{\rm sys}$~[day]     & $180.4(9.8)$   &                 &             \\
$\sigma_{\rm jitter}$~[m/s]& $5.89(56)$     & $3.91(67)$      & $2.93(37)$  \\
r.m.s.~[m/s]               & $7.33$         & $4.00$          & $3.05$      \\
\hline
\multicolumn{4}{c}{general characteristics of the fit} \\
$\tilde l$~[m/s]           & \multicolumn{3}{c}{$5.93$} \\
$d$                        & \multicolumn{3}{c}{$21-4=17$}   \\
\hline
\end{tabular}\\
The orbital inclination of the coplanar system is fixed to $27^\circ$, which is the debris
disk inclination according to \citet{Kennedy13}. See also the notes of Table~\ref{tab_bc}
and Table~\ref{tab_bc_ACR}.
\end{table}

In Table~\ref{tab_bcdY_inc_ACR} we give a refined fit with an ACR constraint and assuming
the value $i=27^\circ$, which is now more likely than $i=90^\circ$, due to
\citet{Kennedy13}. This configuration is stable for at least $10^6$~yrs. The
likelihood-ratio separation between this new fit and the relevant unconstrained fit (a
modification of Table~\ref{tab_bcdY} with free $i$) corresponds to only $1.6\sigma$
(asymptotically). The actual significance is probably even slightly smaller (see
Sect.~\ref{sec_valid}), so we may conclude that the RV data are entirely consistent with
the coplanar three-planet ACR(b,c) configuration inclined by $27^\circ$ to the sky plane.
In fact, such change of $i$ had only a negligible effect on the fit, except for the
absolute planetary masses. Therefore, the value of $i$ does not significantly affect our
main results presented so far, including e.g. the conclusions about the $1100$-day
periodicity. Besides, most of these results were anyway obtained assuming a free-floating
$i$, which was typically located in the range $30^\circ-50^\circ$, and this is not too far
from the value provided by \citet{Kennedy13}.

However, the value of $i=27^\circ$ roughly doubles the planetary mass estimations, and
this may have a significant effect on the planetary dynamics. In particular, the stability
domains around the nominal configuration should significantly shrink. Therefore, we still
need to investigate this effect.

\section{Three-planet dynamics}
\label{sec_dyn}
If the $1100$-day variation is interpreted as a third planet, the entire system appears
remarkably close to a 1:2:5 three-planet resonance. Although the nominal fits presented
above infer that the third planet is still slightly out of the 5:2 resonance, showing a
circulation of critical angles rather than a libration, the parameters of the third planet
are still rather uncertain.

In particular, the eccentricity $e_d$ (along with the pericenter argument $\omega_d$)
looks ill-determined: allowing it to float during the fit generates misleading overfit
effects, like multiple local maxima of the likelihood function. Similarly to the HD37124~c
case discussed in \citet{Baluev08c} or to the GJ876~e case from \citet{Baluev11}, we find
multiple local optima for the eccentric parameters $(e_d,\omega_d)$ at rather high
$e_d\sim 0.3-0.4$. However, all of them are likely unreliable due to the RV model
non-linearity and large uncertainties. All these local solutions may eventually disappear
with more RV observations, as it expectedly occurred in the mentioned HD37124 case
\citep{Wright11}. Moreover, this effect of eccentricity bias looks frequent for exoplanets
discovered by Doppler technique \citep{Beauge12}, and it follows from simulations of the
Keplerian fits done by \citet{Cumming04,ZechKur09} that RV noise favours to large (and
thus overestimated) eccentricity estimations. In the particular case of HD~82943~\emph{d}
this eccentricity bias can be also induced by the correlated astrophysical RV noise.
Therefore, we conclude that the most resonable course of action is to fix $e_d$ at zero.
The actual value of this eccentricity is in fact unconstrained.

The orbital inclination of the system, $i$, is also ill-determined and always remains
statistically consistent with $90^\circ$, although allowing it to float during the fit
does not generate any statistical degeneracies or other obvious bad effects (e.g., the
simulations of Sect.~\ref{sec_valid} were all done with a floating $i$). It follows from
the results by \citet{Kennedy13} that we must pay a particular attention to the value of
$i=27^\circ$.

The orbital period $P_d$ has rather good estimation accuracy, but nevertheless it may be
inside as well as slightly out of the 5:2 resonance, implying a significant change in
the planetary dynamics. Therefore, the uncertainties of $e_d$, $\omega_d$, $P_d$, and $i$
allow for a wide spread of possible dynamical regimes of the three-planet system.

\begin{figure*}
\includegraphics[width=0.49\linewidth]{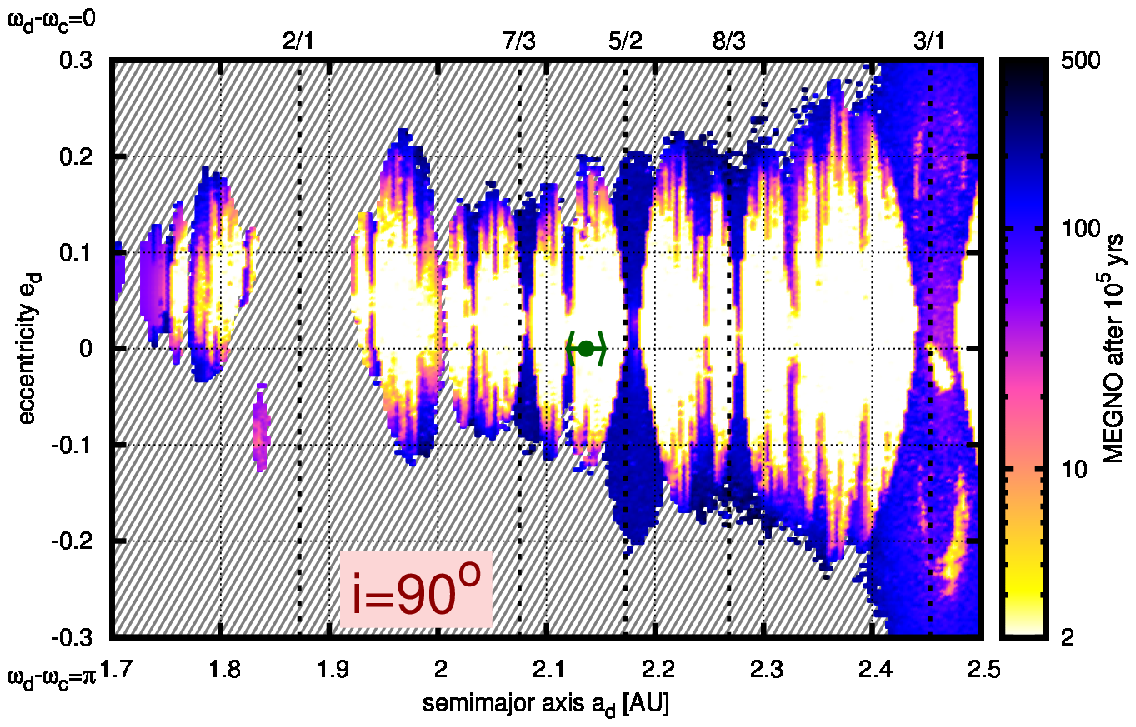}
\includegraphics[width=0.49\linewidth]{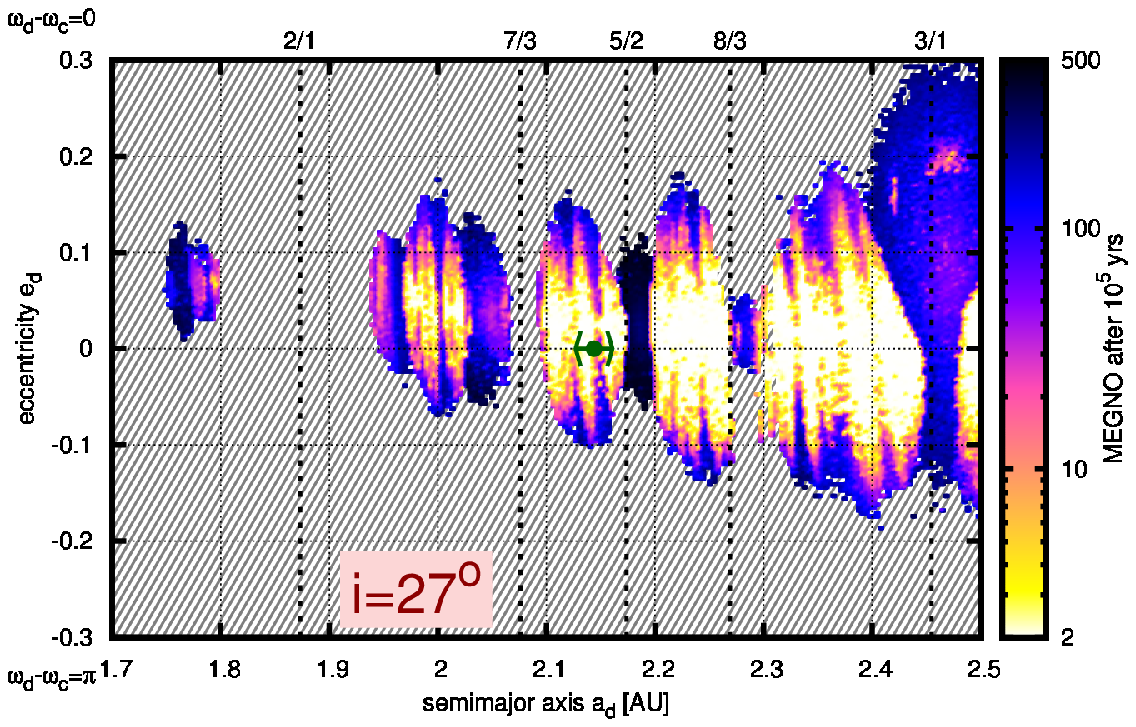}
\caption{Dynamical maps for the neighbourhood of the planet \emph{d} location (marked as a
point with a error bar). The maps are obtained from the fits of Tables~\ref{tab_bcdY_ACR}
(left) and~\ref{tab_bcdY_inc_ACR} (right) by varying $e_d$ and $P_d$. The value of
$\omega_d$ was set to either $\omega_b$ (upper semi-plane in each panel) or $\omega_b+\pi$
(lower semi-plane). The hashed region corresponds to configurations that did not survive
the integration term of $10^5$~yrs. The MEGNO chaoticity indicator is encoded in colour.
The error bar of the nominal position reflects the uncertainty of the period $P_d$
assuming $e_d=0$. Statistical uncertainties of the eccentricity $e_d$ exceed the ordinate
range, so this eccentricity is only constrained by the stability requirement.}
\label{fig_dyn_Pe}
\end{figure*}

Our first task is to analyse the orbital stability of the region of the phase space
exterior to the two known planets of the system. This will help us to constrain the
location of the third planet.

A detailed analysis of the $(a_d,e_d)$ phase space is shown in Fig.~\ref{fig_dyn_Pe}. Here
we show a dynamical map constructed from the numerical integration of two grids of initial
conditions for the outer planet, based on the ACR configurations of
Table~\ref{tab_bcdY_ACR} and Table~\ref{tab_bcdY_inc_ACR}. Positive (negative) values of
$e_d$ correspond to aligned (anti-aligned) orbits with respect to the planet \emph{b}. All
initial conditions were integrated for $10^5$ years. The colour code shows the values of
the MEGNO chaoticity indicator \citep{CincottaSimo00} attained during the integration
interval, while the hashed domain corresponds to initial conditions that implied planetary
ejections or collisions within this time-span (i.e. unstable systems).

Around $a_d=1.87$~AU we can clearly observe the hashed band of the 2/1 MMR. This however
does not mean that the Laplace resonance 1:2:4 would inevitably lead to instability. The
actual stability also depends on the angular variables that were set to particular values
in Fig.~\ref{fig_dyn_Pe}. The plot also shows evidence of both the 5/2 and 3/1 resonances
for larger semimajor axis. Other commensurabilities are also visible, although not as
strong. The nominal configuration is located between the 7/3 and 5/2 MMRs. The 7/3 MMR
becomes unstable for $i=27^\circ$ (at least for the RV-fitted values of the angles), while
the 5/2 MMR remains stable.

These results indicate that it is not difficult to find a stable configuration for the
third planet and in a good statistical agreement with the RV fits. However, the stability
is rather sensitive to apparently small changes of the system parameters. For example, the
stability domain for the \emph{non-ACR} fit of Table~\ref{tab_bcdY} is significantly
reduced in comparison with what we can see in Fig.~\ref{fig_dyn_Pe}. Actually, the nominal
fit of Table~\ref{tab_bcdY} is even unstable. This indicates that apparently minor changes
in the configuration of the two main planets of the system may dramatically affect the
dynamics of the third planet.

The stability domains may be also reduced by assuming a smaller value for the system
inclination $i$, which increases the actual planet masses. However, we found that the
nominal ACR system remains stable for $i$ as small as $17^\circ$, so this limitation is
not very important. We may note that the change of $i$ to $27^\circ$ remarkably
transformed the structure of individual MMRs in the domain, although their general
structure is still similar. In particular, the nominal solution moved very close to a
high-order 22:9 (or possibly 17:7) MMR between the planets \emph{b} and \emph{d}, which
increased the chaoticity in the entire system.

\begin{figure*}
\includegraphics[width=0.49\linewidth]{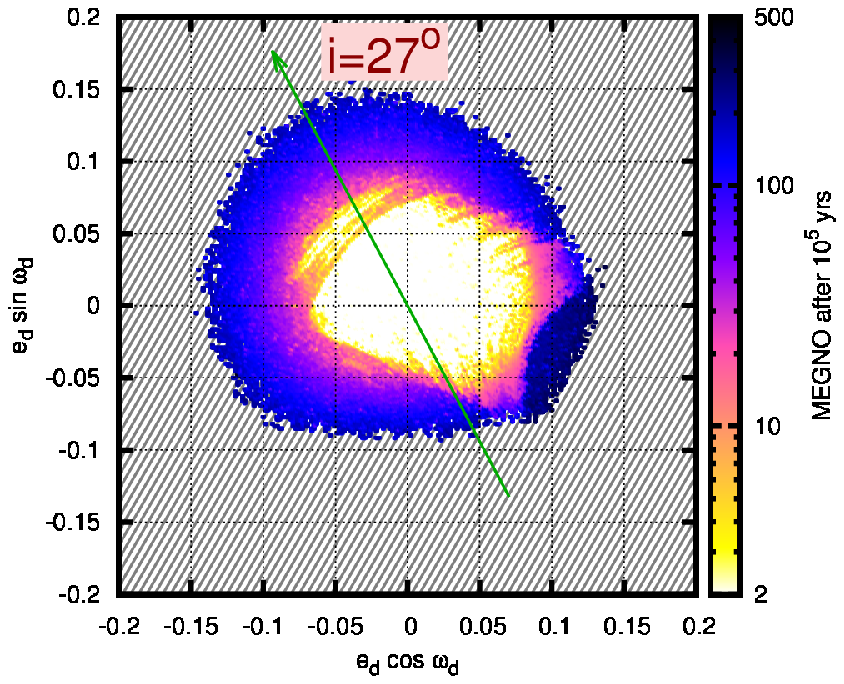}
\includegraphics[width=0.49\linewidth]{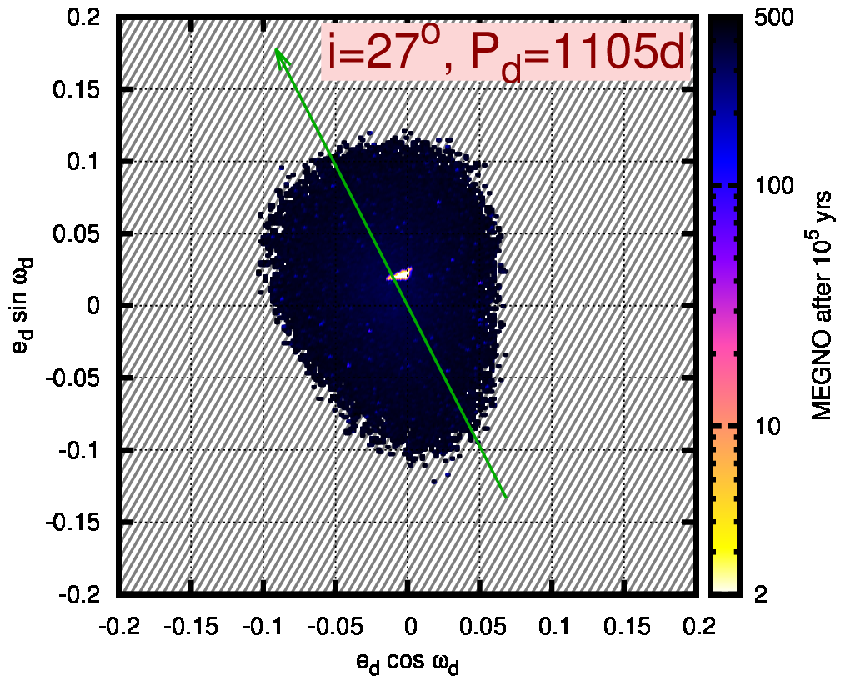}
\caption{Dynamical maps for the ill-determined parameters $e_d$ and $\omega_d$. The left
panel is based on the fit of Table~\ref{tab_bcdY_inc_ACR} with non-resonant best fitting
$P_d=1075$~d, while the right one is based on the fit with $P_d$ fixed at $1105$~d, the
centre of the 5:2 resonance in Fig.~\ref{fig_dyn_Pe}. The arrows set the direction of
$\omega_b$. The other notations are the same as in Fig.~\ref{fig_dyn_Pe}.}
\label{fig_dyn_we}
\end{figure*}

A second pair of dynamical maps is shown in Fig.~\ref{fig_dyn_we}. These are the maps for
the ill-determined eccentric parameters $e_d$ and $\omega_d$ computed for the ACR fit with
$i=27^\circ$ and a free-floating period $P_d$ (eventually estimated by a non-resonant
value), and for the ACR fit assuming the same $i=27^\circ$ and fixing $P_d=1105$~d (at the
resonance 5/2 with the planet \emph{b}). We can see that the stability is generally
favoured by a small $e_d$, although the upper limit on $e_d$ depends on the orientation
angle $\omega_d$. The shape of the stability domain is different for the resonant and
non-resonant value of $P_d$. In the non-resonant case we see clear influence of the 22:9
MMR (e.g. the chaoticity fibers in the left-top part of the stability domain). This
resonance is not very strong, so the major part of the domain does not look affected by
it. Nevertheless, from Fig.~\ref{fig_dyn_Pe} we can see that this resonance could become a
dominating factor after a small increase of $P_d$ with respect to the nominal value. For
the 5:2 MMR case, the chaoticity is much larger, and the relevant stability domain looks
featureless except for a tiny spot of regular motion near the centre. This island of
regular motion appears not belonging to the three-planet MMR; below we discuss this in
more details.

For the triple-resonance case, the angle $\omega_d$ does not describe the secular dynamics
of the third planet comprehensively. In this case, the longitude $\lambda_d$ should also
be considered, since we cannot directly average the resonant Hamiltonian over it. Although
the value of $\lambda_d$ is determined relatively well (at least, with a much better
accuracy than $\omega_d$), we must plot a dynamical map by varying this longitude to
understand the position of the nominal system in the phase space.

\begin{figure}
\includegraphics[width=84mm]{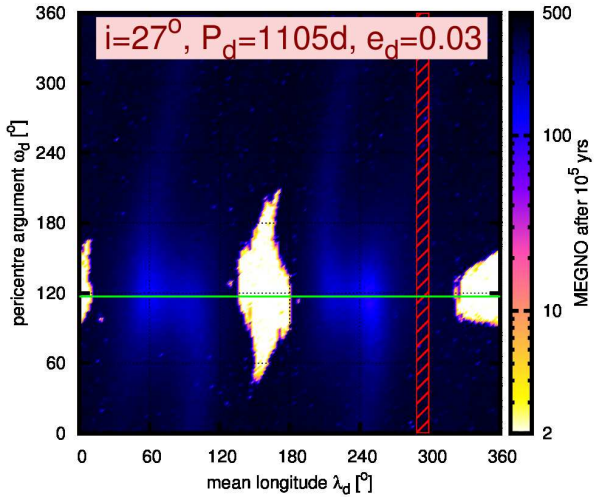}
\caption{Dynamical map for the angular parameters $\lambda_d$ and $\omega_d$ for the
three-planet MMR. The maps are based on the same fit as the one used for the right frame
of Fig.~\ref{fig_dyn_we}, now setting $e_d=0.03$. The horizontal (green) line shows the
position of $\omega_b$. The hashed vertical band shows the one-sigma uncertainty range of
the best fitting $\lambda_d$ (assuming that $\omega_d$ is indetermined). The other
notations are the same as in Fig.~\ref{fig_dyn_Pe}.}
\label{fig_dyn_lw}
\end{figure}

If Fig.~\ref{fig_dyn_lw} we show the map plotted for the parameters $\lambda_d$ and
$\omega_d$. The map was based on an ACR(b,c) fit fixing $i=27^\circ$, $P_d=1105$~d,
$e_d=0$, and with $e_d$ manually moved (without further refitting) from $0$ to $0.03$. In
this plane most of the initial conditions lead to very chaotic motion, although still
stable within the time-span covered by our integration. Our attention is mainly attracted
by two remarkable spots of regular motion near $\lambda_d=150^\circ$ and $350^\circ$. The
detailed investigation showed that this regular motion is not truly resonant: one or both
critical angles, related to planet \emph{d}, circulate. Only the 2:1 resonance is
preserved here, while the 5:2 one is broken. Therefore, these spots represent some
breaches in the structure of the three-planet MMR. We cannot tell anything clear about the
topology of these breaches in the phase space. It is an open question, whether they
represent some disconnected inner caves or they look like pipes passing through the MMR
domain. In the remaining (truly resonant) part of the map there are a few small domains
with a smaller degree of chaoticity (those having a bit lighter colour). The typical
Lyapunov time over the map is only $\sim 250$~yr, but it rises to $\sim 1000$~yr in these
domains. These domains are probably related to some triple-ACR configurations like the one
appearing in Sect.~\ref{sec_migr} below. However, the value of $\lambda_d$ suggested by
the RV data is located \emph{between} these domains, where the chaoticity is high. As the
uncertainty of $\lambda_d$ is rather small, it appears that our RV fits are inconsistent
with these moderately-chaotic domains.

The simultaneous presence of fully resonant (1:2:5) as well as only partly resonant (1:2)
configurations in Fig.~\ref{fig_dyn_lw} indicates that this map represents a slice of the
phase space taken close to the relevant separatrix. This explains why most of these
initial conditions are very chaotic~--- the chaos is typically located near a separatrix.
We believe that the chaoticity may be reduced by seeking a suitable adjustment of the
orbital elements of the two inner planets. The ACR constraint used to obtain the above
fits neglected the perturbalitions from the third planet. Taking them into account would
slightly shift the estimated ACR equilibria. This would not significantly affect the
quality of the RV fit, but the long-term dynamics might change dramatically.

So far, we were unable to find in the vicinity of the nominal fit any regular or at least
low-chaotic motion simultaneously belonging to the three-planet MMR. All our
configurations with regular dynamics are not entirely resonant. We however did not try to
vary the parameters of the two inner planets, which may have a significant effect on the
dynamics of the outermost one. Besides, some real planetary systems do show a chaotic
dynamics (see e.g. the GJ~876 case discussion by \citealt{Marti13}), so we should not
assume that the dynamics of the HD~82943 planetary system have to be regular. We only need
it to be long-term stable.

\section{Three-planet migration}
\label{sec_migr}
The primary goal of this section is to demonstrate that the 1:2:5 three-planet resonance
can be naturally established via the mechanism of the planetary migration.

\begin{figure*}
\includegraphics[width=0.95\linewidth]{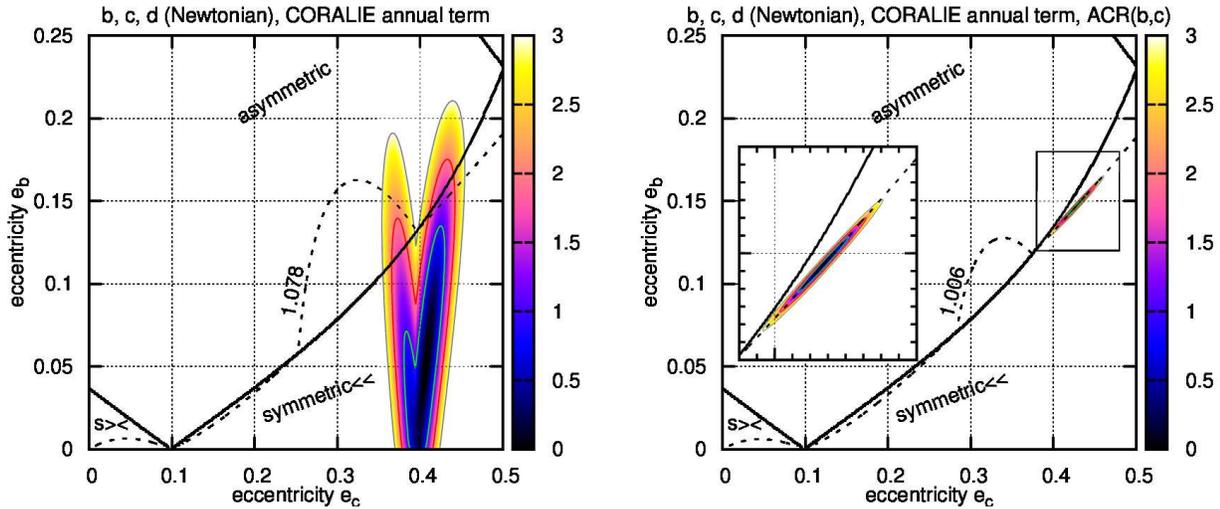}
\caption{The eccentricities of the major planets in the HD82943 system: comparing the
non-ACR (left) and the ACR solution (right). The asymptotic $1$-, $2$-, $3$-sigma
confidence regions inferred by the corresponding orbital model, and the surrounding layout
of the ACR configurations are displayed together. The thick solid lines separate different
types of ACRs for the 2/1 MMR case. The thick dashed line in each graph shows an
iso-family of the ACRs with a constant planet mass ratio, taken from the corresponding RV
fit.}
\label{fig_e12}
\end{figure*}

But first of all, let us investigate in more detail the dynamical status of the two main
resonant planets \emph{c} and \emph{b}. We compare the general layout of the dynamical ACR
families of the 2:1 resonance \citep{Beauge03} with the actual best-fitting configurations
and with the associated parametric uncertainty regions in the plane $(e_c,e_b)$. These
results are plotted in Fig.~\ref{fig_e12}, where we use a three-planet coplanar model with
a free-floating orbital inclination. Each point in the $(e_c,e_b)$ plain corresponds to
some ACR configuration. We have three domains, corresponding to different types of stable
ACRs: the symmetric anti-aligned family (labelled as ``s$><$''), the symmetric aligned one
(``symmetric$<<$''), and the asymmetric ACRs.

Each ACR configuration implies, in particular, a fixed value of the planetary mass ratio
that must be held. Thus, for a given mass ratio we can plot a corresponding iso-family of
ACRs. Such isolines are very important, because they may serve as evolutionary tracks of
the system during the planet migration phase \citep{Beauge06}. In each of the two panels
of Fig.~\ref{fig_e12} we plot a single such isoline that corresponds to the relevant best
fitting mass ratio. We can see that both these isolines pass through all three types of
ACRs. Therefore, this system could undergo an asymmetric ACR state in some past and then
switched back to the symmetric regime.

Planetary migration was simulated using a standard $N$-body code based on a Bulirsch-Stoer
integration routine, plus a Stokes-type exterior force \citep{Beauge06} with specified
values for the $e$-folding times for the semimajor axis ($\tau_a$) and eccentricity
($\tau_e$). We assumed that only the exterior (hypothetical) planet suffered the
migration. Both inner planets suffered no orbital decay, except the indirect one, induced
by the outer planet once the 3-planet resonance was established. However, we did include
an eccentricity damping on the \emph{c} and \emph{b} planets, just to keep their
eccentricities fixed.

\begin{figure}
\includegraphics*[width=84mm]{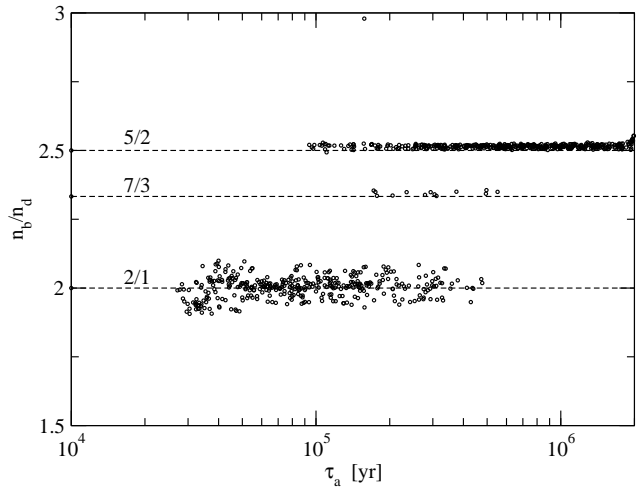}
\caption{Final mean-motion ratios between the two outer planets $m_b$ and $m_d$, as a
function of the orbital decay $e$-folding time, for a total of $1798$ $N$-body runs with
planetary migration acting on the outer mass. The main MMRs attained by the planets are
shown in horizontal dashed lines.}
\label{fig_migres}
\end{figure}

We analysed several initial conditions and migration rates. There is evidence
\citep[e.g.][]{Beauge08,Marti13} that 3-planet resonances may be fairly fragile with small
stability domains, so it is possible that the probability of finding stable configurations
is not high. On the other hand, it is well known that the commensurability in which the
bodies are ultimately captured depend on the migration rate and initial semimajor axis
ratio \citep{NelsonPap02,Rein10}.

The initial orbits of $m_c$ and $m_b$ were chosen equal to those shown in
Table~\ref{tab_bcdY_ACR}, while the semimajor axis and eccentricity of the outer planet
were chosen randomly in the intervals $a_d \in [2.36,2.38]$ AU and $e_d \in [0,0.02]$.
Although the limits of both intervals were small, they guaranteed a random distribution of
the angular variables at each resonance, allowing us to estimate the capture probabilities
in each commensurability. The values of $a_d$ are interior to the 3/1 resonance, but
outside the 5/2. Finally, the migration rates were also chosen randomly in the interval
$\tau_a \in [10^3, 2 \times 10^6]$, while $\tau_e$ was chosen such that $\tau_e/\tau_a =
100$, a value expected for Type-1 migration in laminar isothermal disks
\citep{OgiharaIda09}.

Fig.~\ref{fig_migres} shows the final $P_d/P_b$ ratio for $1798$ initial conditions and
migration rates. The integration time was a function of $\tau_a$, and the runs were
stopped once a stable configuration was reached with constant mean-motion ratios. In all
cases we checked that the two inner planets remained locked in the 2/1 MMR.

As expected, the final MMR attained by the outer planet depends on $\tau_a$. We can
roughly identify four different intervals. For very slow migration rates ($\tau_a > 5
\times 10^5$ yrs), practically all the fictitious systems ended trapped in 3-planet
resonances, and in all cases the two outer planets were locked in the 5/2 MMR. For
slightly faster migrations (down to $\tau_a \sim 10^5$~yrs), some initial conditions
crossed the 5/2 resonance, most being trapped in the 2/1. However, a few were also
captured in the 7/3, while $\sim 10$ per cent of the initial conditions lead to unstable
orbits and were ejected from the system.

\begin{figure}
\includegraphics[width=84mm,clip]{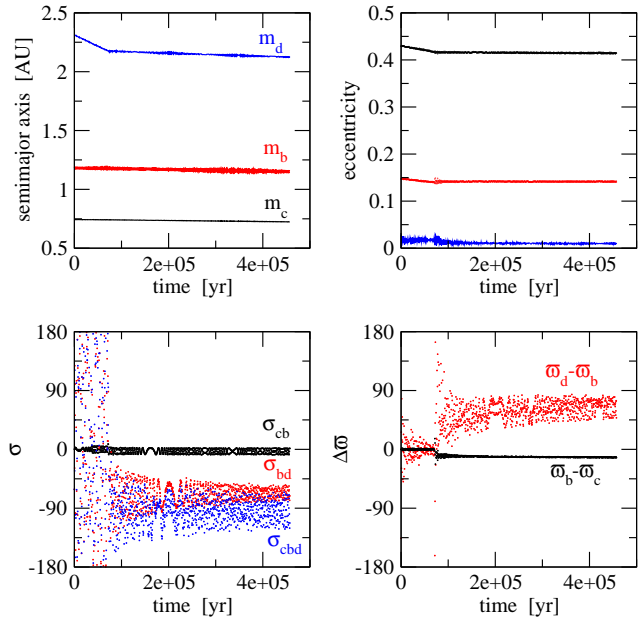}
\caption{A simulation of the planet \emph{d} migration and trapping of all three bodies in
the 5:2:1 multiple resonance. The resonant angles are defined in~(\ref{resang}).}
\label{fig_3trap}
\end{figure}

For even lower decay times ($\tau_a < 10^5$ yrs), the 5/2 resonance seemed unable to
counteract the dissipative force and all stable configurations correspond to the 2/1 MMR,
defining thus a Laplace resonance between all three planets. However, about half of the
runs lead to unstable orbits. Finally, no resonance capture was observed for $\tau_a < 3
\times 10^4$ yrs.

Summarising, the 5:2:1 three-planet resonance seems a natural outcome of this type of
simple $N$-body experiments, as long as the orbital decay rate is sufficiently slow.
Whether this could indeed be the case is a matter of dispute. From the analytical
estimates by \citet{Tanaka02}, for a minimum mass solar nebula and typical disk
properties, $\tau_a$ is estimated to be of the order of $\sim 10^4$~yrs for planets with
masses comparable to the estimated value of $m_d$. However, it is important to keep in
mind that planetary migration is not well understood and it is believed that migration
rates, especially for Type-I, should have been lower than what linear theories for laminar
disks predict. For example, MHD turbulence could delay the orbital decay as much as two
orders of magnitude \citep{NelsonPap04,Alibert05}, leading to values more compatible with
planetary formation. So, the values $\tau_a \sim 10^5$~yrs or even higher are plausible.

Figure \ref{fig_3trap} shows an example of the trapping of third planet in a 5/2 MMR with
the second one, and the consequent locking of all three planets in the 5:2:1 multiple
resonance. The two top frames show the time variation of the semimajor axes (left) and
eccentricities (right). The triple commensurability is attained in less than $10^5$~yrs,
after which all planets continue to migrate together.

The bottom left-hand graph shows the evolution of the resonant angles, defined as:
\begin{eqnarray}
\sigma_{cb} &=& 2 \lambda_b -  \lambda_c - \varpi_c, \nonumber\\
\sigma_{bd} &=& 5 \lambda_d - 2\lambda_b - 3\varpi_b, \nonumber\\
\sigma_{cbd} &=& 5 \lambda_d - 8\lambda_b + 3\lambda_c.
\label{resang}
\end{eqnarray}
where $\sigma_{cbd}$ is the critical angle of the 3-planet resonance. The angle
$\sigma_{cb}$ is the leading critical angle of the 2/1 MMR between the planets \emph{c}
and \emph{b}, and it is always librating. Before the triple-resonance is established,
$\sigma_{cb}$ librates around zero, as expected from a symmetric ACR solution. However,
after the third planet becomes resonant, the libration centre switches to an asymmetric
value, although still close to zero. The angle $\sigma_{bd}$ is the one associated to the
5/2 resonance between \emph{b} and \emph{d}. It circulates before the resonance trapping,
and librates around an asymmetric value after that. The same is noted in $\sigma_{cbd}$.
Note that from e.g. the fit of Table~\ref{tab_bcdY_ACR} we have $\sigma_{cbd} = -9^\circ
\pm 29^\circ$ and $\sigma_{cbd} = -34^\circ \pm 29^\circ$ for
Table~\ref{tab_bcdY_inc_ACR}, so the real orbital configuration is not necessarily a
three-planet ACR like the one appearing in Fig.~\ref{fig_3trap}.

Finally, the bottom right-hand plot shows the behaviour of the difference in pericenters.
Again, $\varpi_b-\varpi_c$ starts librating around zero, but changes to an asymmetric
libration after the system is trapped in the 5:2:1 MMR. The same is also noted for
$\varpi_d-\varpi_b$.

\begin{figure}
\includegraphics*[width=84mm]{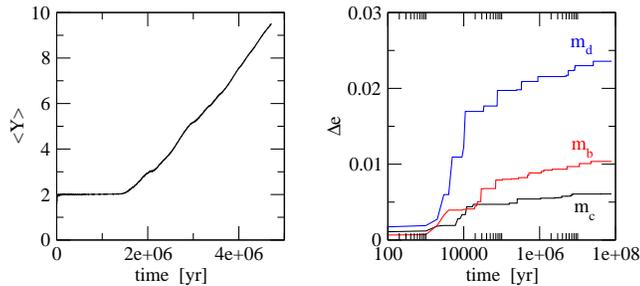}
\caption{Left: MEGNO value $\langle Y \rangle$ as function of time for an initial
condition in the 5:2:1 MMR. Right: Evolution of $\Delta e$ for the three planets during a
$10^8$~yrs $N$-body integration.}
\label{fig_megno}
\end{figure}

To check whether the stability of this orbit is due to the gas, we took the state of the
system at $t = 2 \times 10^5$ yrs, considered those as new initial conditions and
integrated it again without gas effects for $10^8$ yrs. We also calculated the MEGNO chaos
indicator for the run. Results are shown in the left plot of Fig.~\ref{fig_megno}. The
value of $\langle Y \rangle$ starts close to $2$ (indicating a regular motion) but after
$\sim 10^6$~yrs it begins to linearly grow. This indicates chaotic motion, although only
noticeable after $10^6$~yrs, indicating that the chaos is weak. A linear fit of the value
of $\langle Y \rangle$ after this time indicates a maximum Lyapunov exponent also of the
order of $10^6$~yrs.

To analyse the stability of this chaotic configuration, we tried to estimate any diffusion
in the action space. This was done calculating the evolution of $\Delta e$, defined as the
amplitude of the eccentricity of each planet. For regular motion, $\Delta e$ should be
constant. For chaotic but stable motion, this quantity should also be constant or bounded.
Results are shown in the right-hand plot of Fig.~\ref{fig_megno}. We observe a secular
increase in all values of $\Delta e$, larger for the outer planet and smaller for the
inner body. The values, however, remain small. If there is any orbital instability, it
should only be observable for timescales much larger than the age of the star.

\section{Conclusions}
We believe that the existence of the $1100$-days variation in the RV data for HD82943 is
if not convincing, at least plausible. Moreover, it is likely that this variation was
caused by some agent related to the star itself rather than to a particular instrument.
However, we reiterate that we still do not insist on the planetary interpretation of this
variation. It can also be a hint of some long-term noise caused by the star's activity.

The planetary interpretation leads us to an extremely interesting dynamical system in the
three-planet resonance 1:2:5. This would be, to our concern, the second such candidate
system. The previous one was the system of KOI~806/Kepler-30, detected by transits and
transit timing variations \citep{Tingley11}, although later data suggested that its third
planet is significantly out of the $5:2$ MMR \citep{Fabrycky12}.

What concerns the major planets \emph{b} and \emph{c}, their dynamics is likely close to
the aligned ACR located near the border with asymmetric ACRs. But the RV fitting
uncertainties still do not constrain the dynamical regime of HD 82943~\emph{d} well. The
third planet may be inside as well as slightly out of the 5:2 MMR with the planet
\emph{b}, implying different dynamics. Initial conditions with non-resonant third planet
often lead to a regular and stable motion. However, inside the three-planet 1:2:5
resonance, we could not find any regular or low-chaotic motion that would be more or less
consistent with the RV data. It is nonetheless known that chaotic configurations are not
necessarily inacceptable, since chaos does not necessarily imply instability. For example,
the GJ~876 planetary system demonstrates a chaotic but stable motion in the Laplace
resonance \citep{Marti13}.

We find that the three-planet 1:2:5 resonance may represent a rather natural outcome of
the planetary orbital migration. If this three-planet resonance will be further confirmed,
this may place significant constraints on the parameters of the migration process, like
the characteristic migration rate.

Whether the third planet exists or not, there is one interesting matter concerning the two
inner planets. Their nominal configuration corresponds to a symmetric aligned ACR located
very close to the boundary with the domain of asymmetric ACRs. Moreover, from
Fig.~\ref{fig_e12} we may suspect that this system could have passed through the
asymmetric corotation regime somewhen in the past, during the planetary migration stage.

According to \citet{Beauge06}, once the migration is driven by a dissipative adiabatic
force, it should follow the isolines of the constant mass ratio shown in
Fig.~\ref{fig_e12}. Therefore, during the migration there could be two rather abrupt
switches between the symmetric and asymmetric corotation modes. Due to large
eccentricities and masses of the planets \emph{c} and \emph{b}, these bifurcations would
basically represent a dynamical catastrophe for other planets in the system, should they
exist there in that epoch. As a result, some of these planets could be ejected out of the
system or could fall on the star. Such a conclusion provides a nice theoretical
explanation of the spectroscopic observations that detected an unusually high Lithium-6
abundance in the atmosphere of HD82943 \citep{Israelian01}. This chemical anomaly was
hypothetically interpreted as an evidence that some planets could fall on the host star in
the past, enriching it with Lithium-6. We can see that the ACR-sticky migration mechanism
by \citet{Beauge06} provides a good explanation of how such a catastrophe could be
actually triggered.

\section*{Acknowledgements}
This work was supported by the Russian Foundation for Basic Research (project No.
12-02-31119 mol\_a) and by the programme of the Presidium of Russian Academy of Sciences
``Non-stationary phenomena in the objects of the Universe''. This work was also financed
by the Argentinian Research Council -CONICET- and the Universidad Nacional de C\'ordoba
-UNC-. We are grateful to the anonymous reviewer for careful reading and commenting of our
manuscript.

\bibliographystyle{mn2e}
\bibliography{HD82943}

\begin{thebibliography}{}

\bibitem[\protect\citeauthoryear{Alibert, Mordasini, Benz \&
  Winisdoerffer}{Alibert et~al.}{2005}]{Alibert05}
Alibert Y.,  Mordasini C.,  Benz W.,    Winisdoerffer C.,  2005, A\&A, 434, 343

\bibitem[\protect\citeauthoryear{Anglada-Escud{\'e} \&
  Tuomi}{Anglada-Escud{\'e} \& Tuomi}{2012}]{Anglada-Escude12}
Anglada-Escud{\'e} G.,  Tuomi M.,  2012, AA, 548, A58

\bibitem[\protect\citeauthoryear{Baluev}{Baluev}{2008a}]{Baluev08a}
Baluev R.~V.,  2008a, MNRAS, 385, 1279

\bibitem[\protect\citeauthoryear{Baluev}{Baluev}{2008b}]{Baluev08c}
Baluev R.~V.,  2008b, Celest. Mech. Dyn. Astron., 102, 297

\bibitem[\protect\citeauthoryear{Baluev}{Baluev}{2009}]{Baluev08b}
Baluev R.~V.,  2009, MNRAS, 393, 969

\bibitem[\protect\citeauthoryear{Baluev}{Baluev}{2011}]{Baluev11}
Baluev R.~V.,  2011, Celest. Mech. Dyn. Astron., 111, 235

\bibitem[\protect\citeauthoryear{Baluev}{Baluev}{2012}]{Baluev12}
Baluev R.~V.,  2012, MNRAS, 422, 2372

\bibitem[\protect\citeauthoryear{Baluev}{Baluev}{2013a}]{Baluev13a}
Baluev R.~V.,  2013a, MNRAS, 429, 2052

\bibitem[\protect\citeauthoryear{Baluev}{Baluev}{2013b}]{Baluev13c}
Baluev R.~V.,  2013b, Astronomy \& Computing, 2, 18

\bibitem[\protect\citeauthoryear{Beaug\'e, Ferraz-Mello \&
  Michtchenko}{Beaug\'e et~al.}{2003}]{Beauge03}
Beaug\'e C.,  Ferraz-Mello S.,    Michtchenko T.~A.,  2003, ApJ, 593, 1124

\bibitem[\protect\citeauthoryear{Beaug{\'e}, Ferraz-Mello \&
  Michtchenko}{Beaug{\'e} et~al.}{2012}]{Beauge12}
Beaug{\'e} C.,  Ferraz-Mello S.,    Michtchenko T.~A.,  2012, Research in
  Astron. Astrophys., 12, 1044

\bibitem[\protect\citeauthoryear{Beaug{\'e}, Giuppone, Ferraz-Mello \&
  Michtchenko}{Beaug{\'e} et~al.}{2008}]{Beauge08}
Beaug{\'e} C.,  Giuppone C.,  Ferraz-Mello S.,    Michtchenko T.~A.,  2008,
  MNRAS, 385, 2151

\bibitem[\protect\citeauthoryear{Beaug\'e, Michtchenko \&
  Ferraz-Mello}{Beaug\'e et~al.}{2006}]{Beauge06}
Beaug\'e C.,  Michtchenko T.~A.,    Ferraz-Mello S.,  2006, MNRAS, 365, 1160

\bibitem[\protect\citeauthoryear{Beichman, Bryden, Rieke, Stansberry, Trilling,
  Stapelfeldt, Werner, Engelbracht, Blaylock, Gordon, Chen, Su \&
  Hines}{Beichman et~al.}{2005}]{Beihmann05}
Beichman C.~A.,  Bryden G.,  Rieke G.~H.,  Stansberry J.~A.,  Trilling D.~E.,
  Stapelfeldt K.~R.,  Werner M.~W.,  Engelbracht C.~W.,  Blaylock M.,  Gordon
  K.~D.,  Chen C.~H.,  Su K. Y.~L.,    Hines D.~C.,  2005, ApJ, 622, 1160

\bibitem[\protect\citeauthoryear{Cincotta \& Sim{\'o}}{Cincotta \&
  Sim{\'o}}{2000}]{CincottaSimo00}
Cincotta P.~M.,  Sim{\'o} C.,  2000, A\&AS, 147, 205

\bibitem[\protect\citeauthoryear{Cumming}{Cumming}{2004}]{Cumming04}
Cumming A.,  2004, MNRAS, 354, 1165

\bibitem[\protect\citeauthoryear{Dumusque, Pepe, Lovis, S{\'e}gransan,
  Sahlmann, Benz, Bouchy, Mayor, Queloz, Santos \& Udry}{Dumusque
  et~al.}{2012}]{Dumusque12}
Dumusque X.,  Pepe F.,  Lovis C.,  S{\'e}gransan D.,  Sahlmann J.,  Benz W.,
  Bouchy F.,  Mayor M.,  Queloz D.,  Santos N.,    Udry S.,  2012, Nature, 491,
  207

\bibitem[\protect\citeauthoryear{Fabrycky, Ford, Steffen, Rowe, Carter,
  Moorhead, Batalha, Borucki, Bryson, Buchhave, Christiansen, Ciardi, Cochran,
  Endl, Fanelli, Fischer, Fressin, Geary, Haas, Hall, Holman, Jenkins, Koch \&
  Latham}{Fabrycky et~al.}{2012}]{Fabrycky12}
Fabrycky D.~C.,  Ford E.~B.,  Steffen J.~H.,  Rowe J.~F.,  Carter J.~A.,
  Moorhead A.~V.,  Batalha N.~M.,  Borucki W.~J.,  Bryson S.,  Buchhave L.~A.,
  Christiansen J.~L.,  Ciardi D.~R.,  Cochran W.~D.,  Endl M.,  Fanelli M.~N.,
  Fischer D.,  Fressin F.,  Geary J.,  Haas M.~R.,  Hall J.~R.,  Holman M.~J.,
  Jenkins J.~M.,  Koch D.~G.,    Latham D.~W.  et al., 2012, ApJ, 750, 114

\bibitem[\protect\citeauthoryear{Ferraz-Mello, Michtchenko \&
  Beaug{\'e}}{Ferraz-Mello et~al.}{2005a}]{FerrazMello05}
Ferraz-Mello S.,  Michtchenko T.~A.,    Beaug{\'e} C.,  2005a, ApJ, 621, 473

\bibitem[\protect\citeauthoryear{Ferraz-Mello, Michtchenko, Beaug{\'e} \&
  Callegari}{Ferraz-Mello et~al.}{2005b}]{Ferraz-Mello-lec1}
Ferraz-Mello S.,  Michtchenko T.~A.,  Beaug{\'e} C.,    Callegari N.,  2005b,
  Lect. Not. Phys., 683, 219

\bibitem[\protect\citeauthoryear{Go{\'z}dziewski, Breiter \&
  Borczyk}{Go{\'z}dziewski et~al.}{2008}]{Gozd08}
Go{\'z}dziewski K.,  Breiter S.,    Borczyk W.,  2008, MNRAS, 383, 989

\bibitem[\protect\citeauthoryear{Go{\'z}dziewski \& Konacki}{Go{\'z}dziewski \&
  Konacki}{2006}]{GozdKon06}
Go{\'z}dziewski K.,  Konacki M.,  2006, ApJ, 647, 573

\bibitem[\protect\citeauthoryear{Go\'{z}dziewski, Konacki \&
  Maciejewski}{Go\'{z}dziewski et~al.}{2005}]{Gozd05}
Go\'{z}dziewski K.,  Konacki M.,    Maciejewski A.~J.,  2005, ApJ, 622, 1136

\bibitem[\protect\citeauthoryear{Go{\'z}dziewski \&
  Maciejewski}{Go{\'z}dziewski \& Maciejewski}{2001}]{GozdMac01}
Go{\'z}dziewski K.,  Maciejewski A.~J.,  2001, ApJ, 563, L81

\bibitem[\protect\citeauthoryear{Israelian, Santos, Mayor \& Rebolo}{Israelian
  et~al.}{2001}]{Israelian01}
Israelian G.,  Santos N.~C.,  Mayor M.,    Rebolo R.,  2001, Nature, 411, 163

\bibitem[\protect\citeauthoryear{Ji, Kinoshita, Liu, Li \& Nakai}{Ji
  et~al.}{2003}]{Ji03}
Ji J.,  Kinoshita H.,  Liu L.,  Li G.,    Nakai H.,  2003, Celest. Mech. Dyn.
  Astron., 87, 113

\bibitem[\protect\citeauthoryear{Kennedy, Wyatt, Bryden, Wittenmyer \&
  Sibthorpe}{Kennedy et~al.}{2013}]{Kennedy13}
Kennedy G.~M.,  Wyatt M.~C.,  Bryden G.,  Wittenmyer R.,    Sibthorpe B.,
  2013, MNRAS, 436, 898

\bibitem[\protect\citeauthoryear{Lee, Butler, Fischer, Marcy \& Vogt}{Lee
  et~al.}{2006}]{Lee06}
Lee M.~H.,  Butler R.~P.,  Fischer D.~A.,  Marcy G.~W.,    Vogt S.~S.,  2006,
  ApJ, 641, 1178

\bibitem[\protect\citeauthoryear{Mart{\'i}, Giuppone \& Beaug{\'e}}{Mart{\'i}
  et~al.}{2013}]{Marti13}
Mart{\'i} J.~G.,  Giuppone C.~A.,    Beaug{\'e} C.,  2013, MNRAS, 433, 928

\bibitem[\protect\citeauthoryear{Mayor, Udry, Naef, Pepe, Queloz, Santos \&
  Burnet}{Mayor et~al.}{2004}]{Mayor04}
Mayor M.,  Udry S.,  Naef D.,  Pepe F.,  Queloz D.,  Santos N.~C.,    Burnet
  M.,  2004, A\&A, 415, 391

\bibitem[\protect\citeauthoryear{Michtchenko, Beaug\'e \&
  Ferraz-Mello}{Michtchenko et~al.}{2006}]{Michtchenko06}
Michtchenko T.~A.,  Beaug\'e C.,    Ferraz-Mello S.,  2006, Celest. Mech. Dyn.
  Astron., 94, 411

\bibitem[\protect\citeauthoryear{Nelson \& Papaloizou}{Nelson \&
  Papaloizou}{2002}]{NelsonPap02}
Nelson R.~P.,  Papaloizou J. C.~B.,  2002, MNRAS, 333, L26

\bibitem[\protect\citeauthoryear{Nelson \& Papaloizou}{Nelson \&
  Papaloizou}{2004}]{NelsonPap04}
Nelson R.~P.,  Papaloizou J. C.~B.,  2004, MNRAS, 350, 849

\bibitem[\protect\citeauthoryear{Ogihara \& Ida}{Ogihara \&
  Ida}{2009}]{OgiharaIda09}
Ogihara M.,  Ida S.,  2009, ApJ, 699, 824

\bibitem[\protect\citeauthoryear{Rein, Papaloizou \& Kley}{Rein
  et~al.}{2010}]{Rein10}
Rein H.,  Papaloizou J. C.~B.,    Kley W.,  2010, A\&A, 510, A4

\bibitem[\protect\citeauthoryear{Tan, Payne, Lee, Ford, Howard, Johnson, Marcy
  \& Wright}{Tan et~al.}{2013}]{Tan13}
Tan X.,  Payne M.~J.,  Lee M.~H.,  Ford E.~B.,  Howard A.~W.,  Johnson J.~A.,
  Marcy G.~W.,    Wright J.~T.,  2013, ApJ, 777, id101

\bibitem[\protect\citeauthoryear{Tanaka, Takeuchi \& Ward}{Tanaka
  et~al.}{2002}]{Tanaka02}
Tanaka H.,  Takeuchi T.,    Ward W.~R.,  2002, ApJ, 565, 1257

\bibitem[\protect\citeauthoryear{Tingley, Palle, Parviainen, Deeg,
  Zapatero~Osorio, Cabrera-Lavers, Belmonte, Rodriguez, Murgas \&
  Ribas}{Tingley et~al.}{2011}]{Tingley11}
Tingley B.,  Palle E.,  Parviainen H.,  Deeg H.~J.,  Zapatero~Osorio M.~R.,
  Cabrera-Lavers A.,  Belmonte J.~A.,  Rodriguez P.~M.,  Murgas F.,    Ribas
  I.,  2011, A\&A, 536, id. L9

\bibitem[\protect\citeauthoryear{Vuong}{Vuong}{1989}]{Vuong89}
Vuong Q.~H.,  1989, Econometrica, 57, 307

\bibitem[\protect\citeauthoryear{Wright, Veras, Ford, Johnson, Marcy, Howard,
  Isaacson, Fischer, Spronck, Anderson \& Valenti}{Wright
  et~al.}{2011}]{Wright11}
Wright J.~T.,  Veras D.,  Ford E.~B.,  Johnson J.~A.,  Marcy G.~W.,  Howard
  A.~W.,  Isaacson H.,  Fischer D.~A.,  Spronck J.,  Anderson J.,    Valenti
  J.,  2011, ApJ, 730, id93

\bibitem[\protect\citeauthoryear{Zechmeister \& K{\"u}rster}{Zechmeister \&
  K{\"u}rster}{2009}]{ZechKur09}
Zechmeister M.,  K{\"u}rster M.,  2009, A\&A, 496, 577

\end{thebibliography}

\bsp

\label{lastpage}

\end{document}